\documentclass[aps,10pt,preprintnumbers]{revtex4-2}
\usepackage{graphicx}
\usepackage{amsmath}

\begin{document}
\preprint{FERMILAB-FN-1152-AD-ND, LA-UR-22-21987}
\title{PIP2-BD: GeV Proton Beam Dump at Fermilab’s PIP-II Linac}


\author{M.~Toups}\thanks{Contact: toups@fnal.gov, vdwater@lanl.gov}
\affiliation{Fermi National Accelerator Laboratory, Batavia, IL 60510, USA}

\author{R.G. Van de Water}\thanks{Contact: toups@fnal.gov, vdwater@lanl.gov}
\affiliation{Los Alamos National Laboratory, Los Alamos, NM 87545, USA}

\author{Brian Batell}
\affiliation{University of Pittsburgh, Pittsburgh, PA 15260, USA}

\author{S.J.~Brice}
\affiliation{Fermi National Accelerator Laboratory, Batavia, IL 60510, USA}

\author{Patrick deNiverville}
\affiliation{Los Alamos National Laboratory, Los Alamos, NM 87545, USA}

\author{Bhaskar Dutta, Aparajitha Karthikeyan, Doojin Kim, Nityasa Mishra, Adrian Thompson}
\affiliation{Mitchell Institute for Fundamental Physics and Astronomy, Department of Physics and Astronomy, Texas A\&M University, College Station, TX 77843, USA}

\author{Jeff Eldred}
\affiliation{Fermi National Accelerator Laboratory, Batavia, IL 60510, USA}

\author{Roni Harnik}
\affiliation{Fermi National Accelerator Laboratory, Batavia, IL 60510, USA}


\author{Kevin J. Kelly}
\affiliation{Fermi National Accelerator Laboratory, Batavia, IL 60510, USA}

\author{Tom Kobilarcik}
\affiliation{Fermi National Accelerator Laboratory, Batavia, IL 60510, USA}

\author{Gordan Krnjaic}
\affiliation{Fermi National Accelerator Laboratory, Batavia, IL 60510, USA}

\author{B. R. Littlejohn}
\affiliation{Illinois Institute of Technology, Chicago, IL 60616, USA}

\author{Bill Louis}
\affiliation{Los Alamos National Laboratory, Los Alamos, NM 87545, USA}

\author{Pedro A.~N. Machado}
\affiliation{Fermi National Accelerator Laboratory, Batavia, IL 60510, USA}

\author{V.~Pandey}
\affiliation{Fermi National Accelerator Laboratory, Batavia, IL 60510, USA}

\author{Z.~Pavlovic}
\affiliation{Fermi National Accelerator Laboratory, Batavia, IL 60510, USA}

\author{William Pellico}
\affiliation{Fermi National Accelerator Laboratory, Batavia, IL 60510, USA}

\author{Michael Shaevitz}
\affiliation{Columbia University, New York, NY 10027, USA}

\author{P. Snopok}
\affiliation{Illinois Institute of Technology, Chicago, IL 60616, USA}

\author{Rex Tayloe}
\affiliation{Indiana University, Bloomington, IN 47405, USA}

\author{R. T. Thornton}
\affiliation{Los Alamos National Laboratory, Los Alamos, NM 87545, USA}

\author{Jacob Zettlemoyer}
\affiliation{Fermi National Accelerator Laboratory, Batavia, IL 60510, USA}

\author{Bob Zwaska}
\affiliation{Fermi National Accelerator Laboratory, Batavia, IL 60510, USA}

\author{Timothy Hapitas}
\affiliation{Department of Physics, Carleton University, Ottawa, Ontario K1S 5B6, Canada}

\author{Douglas Tuckler}
\affiliation{Department of Physics, Carleton University, Ottawa, Ontario K1S 5B6, Canada}

\author{Jaehoon Yu}
\affiliation{Department of Physics, University of Texas, Arlington, TX 76019, USA}

\begin{abstract}
\end{abstract}

\maketitle
\vspace{-4em}
\section{Physics Goals and Motivation}
Two recent developments in particle physics clearly establish the need for a GeV-scale high energy physics (HEP) beam dump facility. First, theoretical work has highlighted not only the viability of sub-GeV dark sectors models to explain the cosmological dark matter (DM) abundance but also that a broad class of these models can be tested with accelerator-based, fixed-target experiments, which complement growing activity in sub-GeV direct DM detection~\cite{Alexander:2016aln, Battaglieri:2017aum, BRN}. Second, the observation of coherent elastic neutrino-nucleus scattering (CEvNS)~\cite{Freedman:1973yd, Kopeliovich:1974mv} by the COHERENT experiment~\cite{COHERENT:2017ipa, COHERENT:2020iec} provides a novel experimental tool that can now be utilized to search for physics beyond the Standard Model (SM) in new ways, including in searches for light DM~\cite{deNiverville:2015mwa} and active-to-sterile neutrino oscillations~\cite{Anderson:2012pn}, which would provide smoking-gun evidence for the existence of sterile neutrinos.\\

The completion of the PIP-II superconducting LINAC at Fermilab as a proton driver for DUNE/LBNF in the late 2020s creates an attractive opportunity to build such a dedicated beam dump facility at Fermilab.  A unique feature of this Fermilab beam dump facility is that it can be optimized from the ground up for HEP. Thus, relative to spallation neutron facilities dedicated to neutron physics and optimized for neutron production operating at a similar proton beam power, a HEP-dedicated beam dump facility would allow for better sensitivity to dark sector models, sterile neutrinos, and CEvNS-based searches for nonstandard neutrino interactions, as well as more precise measurements of neutrino interaction cross sections relevant for supernova neutrino detection. For example, the Fermilab facility could be designed to suppress rather than maximize neutron production and implement a beam dump made from a lighter target such as carbon, which can have a pion-to-proton production ratio up to $\sim$2 times larger than the heavier Hg or W targets used at spallation neutron sources. The facility could also accommodate multiple, 100-ton-scale HEP experiments located at different distances from the beam dump and at different angles with respect to the beam. This flexibility would allow for sensitive dark sector and sterile neutrino searches, which can constrain uncertainties in expected signal and background rates by making relative measurements at different distances and angles.

\newpage
\section{PIP-II Linac}

\begin{figure}
\centering
  \includegraphics[width=0.8\textwidth]{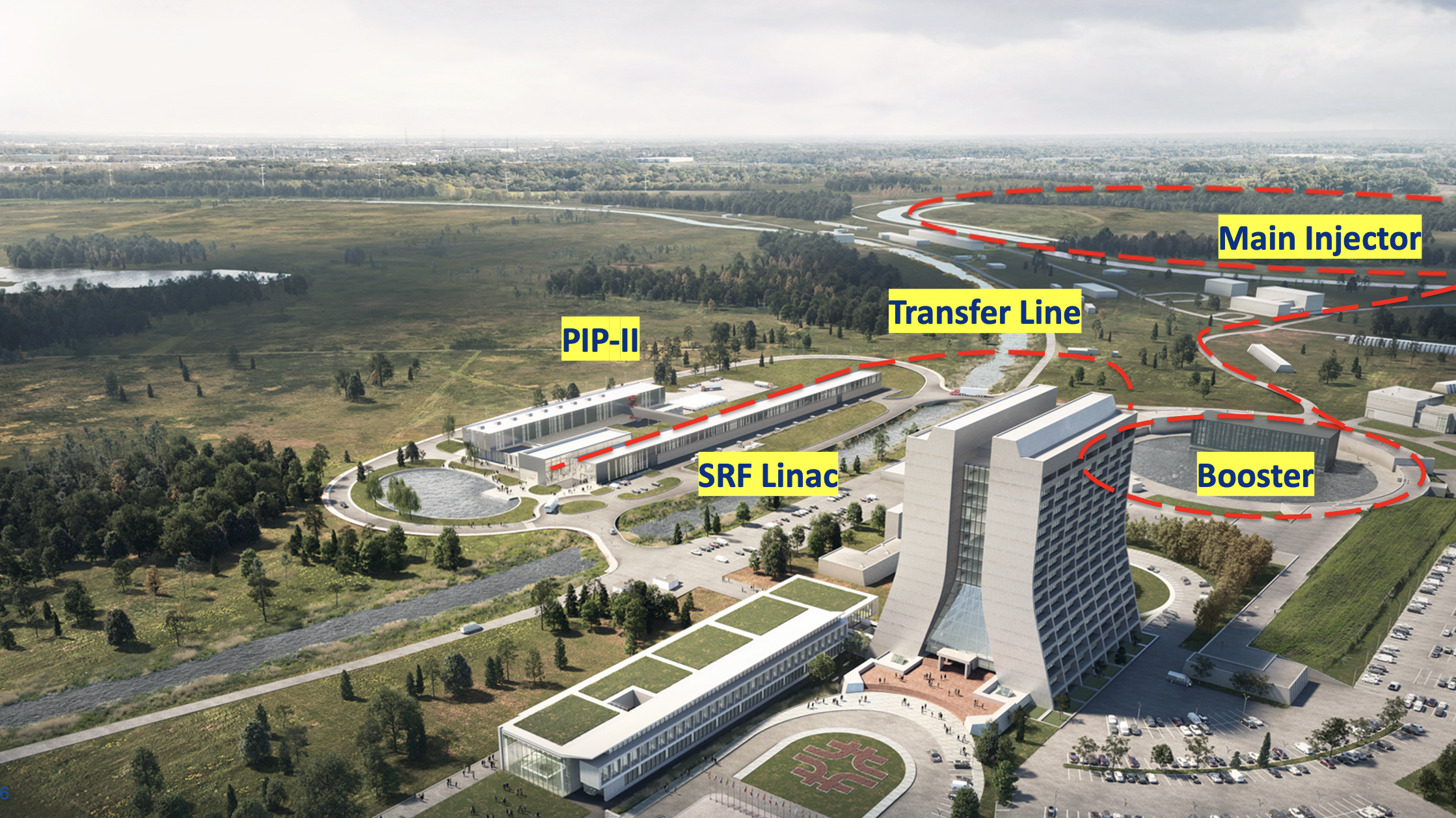}  
  \caption{View of the Fermilab accelerator complex. A rendering of the PIP-II accelerator LINAC is shown in the foreground along with the transfer line to the existing Booster and Main Injector beamlines.}
  \label{fig:PIP2}
\end{figure}

The Proton Improvement Project II (PIP-II) is the first phase of a major transformation of the accelerator complex underway at Fermilab to prepare the lab to host the Deep Underground Neutrino Experiment (DUNE) (see Fig.~\ref{fig:PIP2}). At the heart of this upgrade is the PIP-II superconducting Linac, which serves as the proton driver for the long-baseline neutrino beam to DUNE. The PIP-II linac has a peak current of 2 mA and will initially operate with a duty factor of 1.1\%, accelerating a 0.55 ms pulse of H$^{-}$ ions to 800 MeV and injecting them into the Booster at a rate of 20 Hz~\cite{Lebedev:2017vnu}. This operational mode is sufficient to support an initial 1.2 MW beamline for DUNE at 120 GeV.\\

The PIP-II Linac was designed, however, with the flexibility to support multiple users in mind and is also compatible with continuous wave operation.  Thus, PIP-II is capable of simultaneously supporting a high-power long-baseline neutrino beam and supplying an additional 1.6 MW of proton beam power at 800 MeV.  Furthermore, the PIP-II Linac tunnel includes space and infrastructure to reach 1~GeV and space to add an RF separator for beam sharing to multiple users.

\section{Accumulator Ring Scenarios} \label{sec:accring}

For optimal sensitivity to rare event searches for light DM and short-baseline neutrino anomalies, an accumulator ring is required to bunch the PIP-II beam current into short proton pulses.  Steady-state backgrounds due to cosmic ray activity and internal radiological sources can then be efficiently rejected through beam timing.  Some rejection of beam-related neutron and neutrino backgrounds (in the case of light DM searches) is also possible with very short ($<30$~ns) proton pulses.\\

\begin{figure}
\centering
  \includegraphics[width=0.8\textwidth]{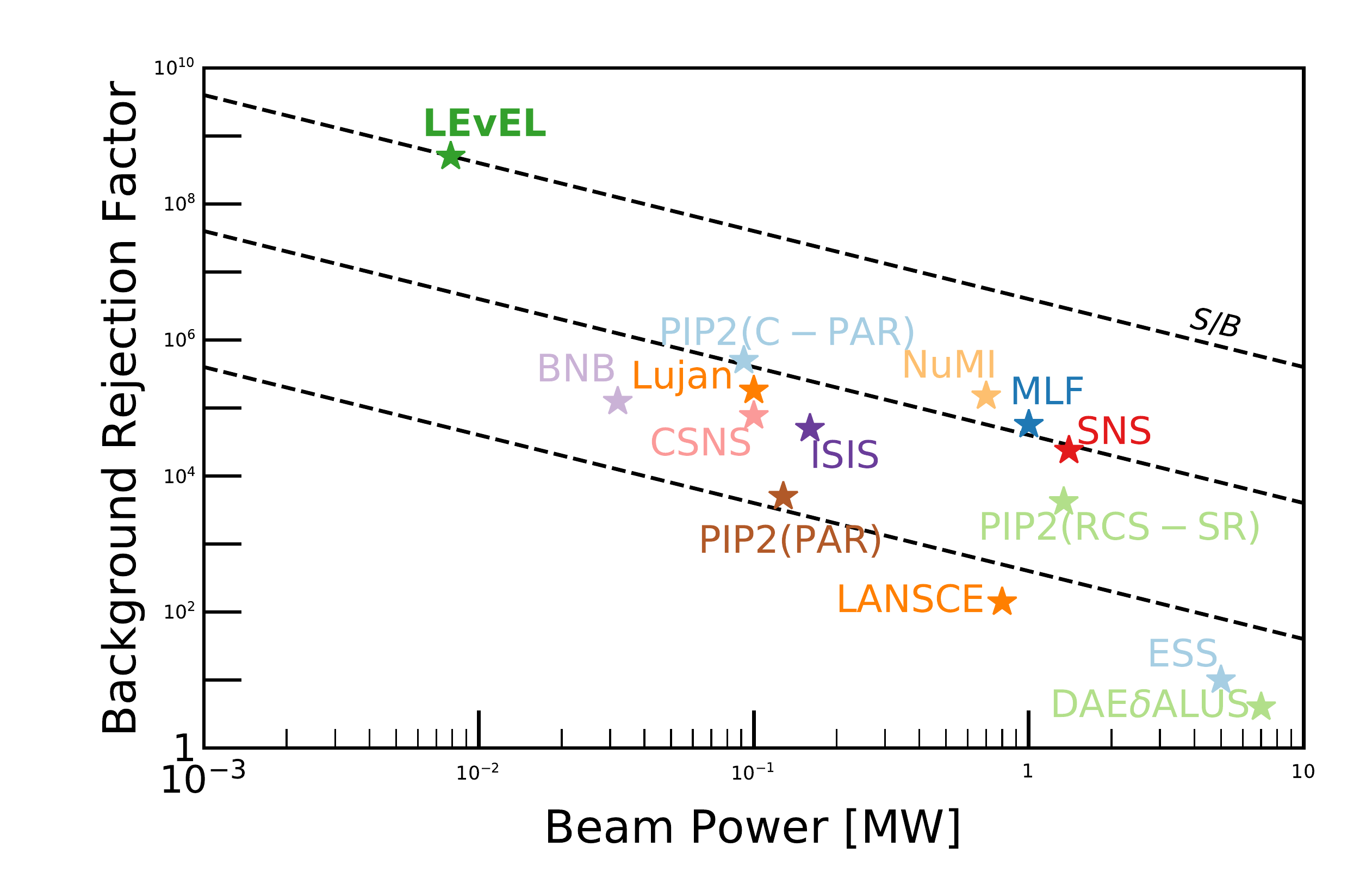}
  \caption{Existing and planned stopped-pion sources across the world along with lines showing a constant signal/background ratio, assuming signal is proportional to beam power and background is inversely proportional to the duty factor. The PIP-II accumulator ring scenarios are shown with the baseline PAR and the possible improvements either increasing the background rejection factor with the C-PAR scenario or increasing the beam power with the RCS-SR scenario. Figure adapted from Ref.~\cite{Kelly:2021jgj}.}
  \label{fig:staging}
\end{figure}
\newpage

We discuss below three different accumulator ring scenarios proposed as upgrades of the Fermilab accelerator complex studied in this paper. These scenarios are staged such that the first scenario, called ``PAR'', could be built within the decade and support an initial beam dump physics program.  The two other accumulator ring scenarios extend the physics reach of PAR either by moving to shorter proton pulses (``C-PAR'') or by increasing the beam power (``RCS-SR'') as shown in Fig.~\ref{fig:staging}. In either case, the PIP2-BD setup would be capable of supporting a world-class physics program. Each accumulator ring is described in detail below with key parameters for each scenario summarized in Table~\ref{table:values}.

\begin{table} [h]
\begin{center}
\begin{tabular}{|p{1.5cm}|p{1.5cm}|p{1.5cm}|p{1.5cm}|p{1.5cm}|}
\hline
Facility & Beam energy (GeV) & Repetition rate (Hz) & Pulse length (s) & Beam power (MW)\\
\hline
PAR & 0.8 & 100 & $2 \times 10^{-6}$ & 0.1 \\
C-PAR & 1.2 & 100 & $2\times 10^{-8}$ & 0.09\\
RCS-SR & 2 & 120 & $2 \times 10^{-6}$ & 1.3 \\
\hline
\end{tabular} \caption{The parameters of three possible accumulator ring scenarios considered to bunch the beam current from the PIP-II linac. See text for more details.}\label{table:values}
\end{center}
\end{table}
\subsection{PAR: PIP-II Accumulator Ring}

The most straightforward design for an accumulator ring would be integrated into the low energy beam transport line between the PIP-II linac and the Booster accelerator ring and would be sited between them (see Ref.~\cite{Pellico}).  This accumulator ring, called ``PAR'', would consist of a pair of interconnected, DC-powered or permanent magnet accumulator rings in the same beam enclosure, each with 1/2 the circumference of the Booster.  PAR could be designed to operate at 800~MeV but with an upgrade path allowing for future operation in the GeV range.  Initially, PAR would provide 100~kW of beam power, limited by stripping foil heating, and have a $\mathcal{O}(10^{-4})$ duty factor. Upgrading the beam energy to 1 GeV would allow for a factor of $\sim$1.3 increase in accumulator ring intensity and implementing laser stripping technology would allow the beam power to increase by another factor of $\sim$4, at the expense of an increase in duty factor by an equivalent factor.

\subsection{C-PAR: Compact PIP-II Accumulator Ring}

An alternative concept for an accumulator ring design that bunches the PIP-II beam into short pulses would enable both a world-class GeV proton beam dump facility and a world-class muon-based charged-lepton flavor violation (CLFV) program~\cite{Prebys,Bernstein}. This would be a green field accumulator ring with a smaller circumference of 100 m.  A design with four 12 m straight sections and the remaining 52 m used for bending will work if the average bending field is 0.8 T (with 1--1.2 T in the dipoles), and the beam energy is no greater than 1.2 GeV.\\

We assume a beam emittance of 24 $\pi$-mm-mrad and 16 RF buckets, each about 25 ns long, such that at 1.2 GeV each bunch would contain $1.2\times 10^{12}$ protons.  With these parameters the space-charge tune shift shouldn't be an issue ($\Delta Q_{SC}=0.22$).  If we fill and extract from the ring at 100 Hz, then we can phase rotate 4 bunches into one 20 ns bunch such that each bunch contains $4.8\times 10^{12}$ protons.\\

In the CLFV-operating mode, C-PAR would fill every other RF bucket (8 out of 16) and extract one bunch at a time to provide $\sim$0.9$~\times~10^{12}$ protons delivered over less than 20~ns. The 20~ns pulses would be consistent with the beam requirements of a CLFV experiment based on the PRISM (Phase Rotated Intense Source of Muons) concept~\cite{Prebys,Bernstein,PRISM}. The 1.2~GeV proton bunches would be extracted at a rate of up to 800~Hz (possibly alternating between several high repetition rate kickers), to deliver 140~kW of beam power to the CLFV experiments. Since the 8 proton bunches are extracted one at at time, the C-PAR ring would be filled by H$^-$ injection at a rate of 100~Hz (similar to the PIP2-BD operating mode). The injected bunch charge would be lower than the PIP2-BD operating mode, but the total ring intensity and beam power would need to be a factor of two higher.

\subsection{RCS-SR: Rapid Cycling Synchrotron Storage Ring}

DUNE ultimately requires an upgraded neutrino beamline reaching 2.4 MW of beam power at 120 GeV. The current Booster ring is incapable of accelerating a sufficient number of protons to reach this beam power and so needs to be replaced by a modern higher-power accelerator.  The Fermilab accelerator complex upgrade scenario laid out in Ref.~\cite{Ainsworth:2021ahm} combines a 2 GeV upgrade of the PIP-II linac with a new 570 m Rapid-Cycling Synchrotron (RCS) replacing the Booster in order to meet the DUNE beamline requirement.\\

This scenario also outlines how a new 2 GeV Storage Ring (SR) built in tandem with the RCS would facilitate H$^-$ injection into the RCS.  This same ring would be capable of delivering intense pulses of $\sim35\times10^{12}$ protons to a new 2 GeV beamline. The SR pulse rate would be constrained by H$^-$ foil-stripping injection, but could be as high as 60 or 120 Hz with H$^-$ laser stripping.


\section{Beam Target Simulation}

PIP2-BD is a platform for rare event searches for light DM and short-baseline neutrino anomalies.  The neutrinos from a stopped-pion decay-at-rest neutrino source are produced via the decays of charged pions and their daughter muons produced by proton collisions with the target. Additionally, many DM models predict DM production via neutral mesons such as pions and etas, also produced by the proton collisions with the target. To model the production of light mesons and other particles within the beam dump, a custom Geant4-based (v4.10.6p01) simulation~\cite{geant1, geant2} implementing a simplified GDML-based target geometry was built to study the accumulator ring scenarios described in Section~\ref{sec:accring}.  The target was modeled as a block of graphite measuring 3.6 m along the beam axis and 27 cm in the directions transverse to the beam. The proton beam was modeled as a Gaussian with a 4.5~cm width and a tuneable energy.\\

\begin{figure}
\centering
  \includegraphics[width=0.7\textwidth]{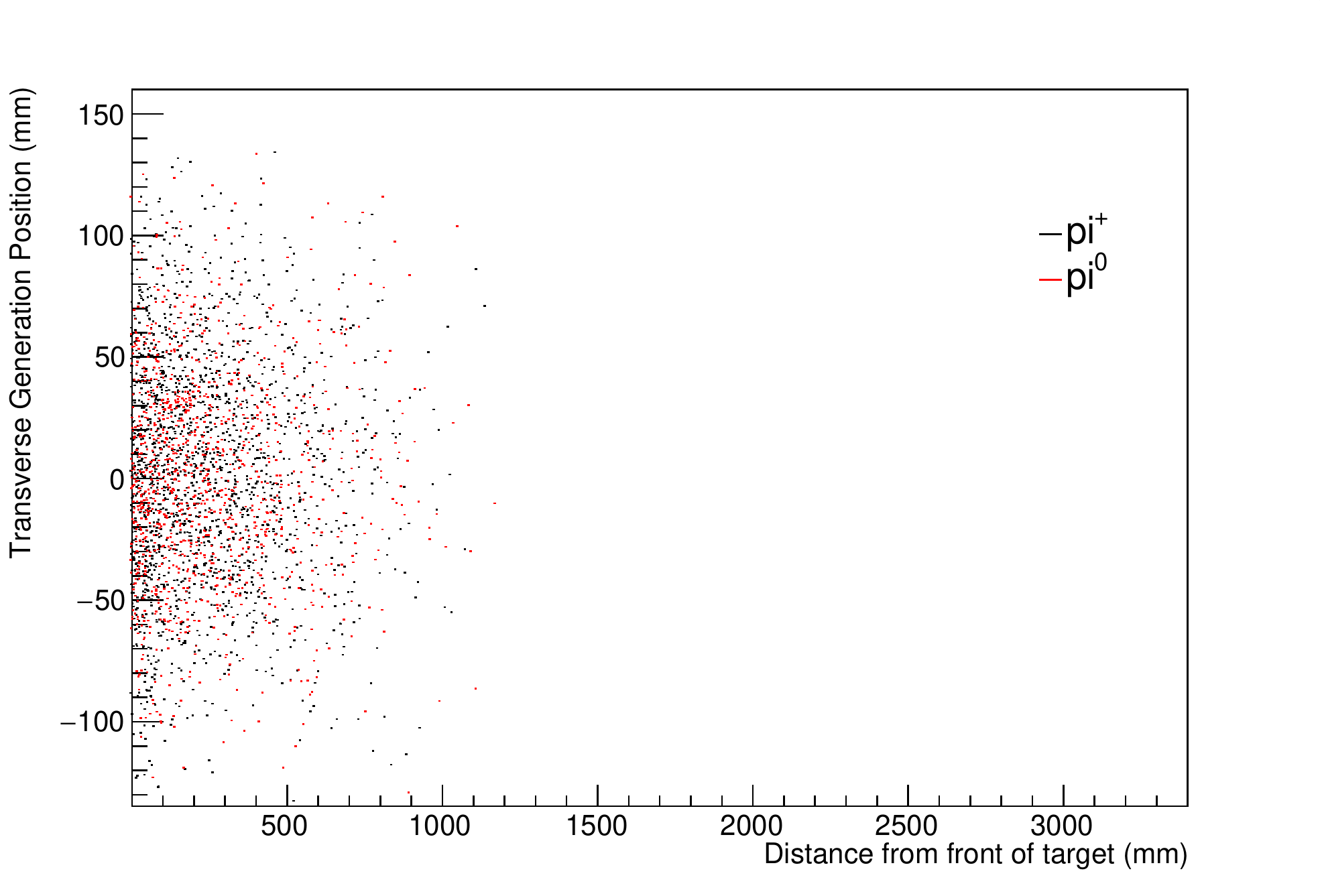}
  \caption{Positions of $\pi^+$ and $\pi^0$ created within the graphite target for both $\pi^+$ and $\pi^0$ considering a simulation of an 800 MeV proton beam using the QGSP\_BERT\_HP hadron production physics list in Geant4.}
  \label{fig:hadronproduction}
\end{figure}

The primary Geant4 physics list used for beam simulations was QGSP\_BERT\_HP, although simulations of eta production used the QGSP\_BIC\_HP physics list.  For an 800 MeV proton beam, the simulation predicts a 21~cm standard deviation for light meson production in the beam direction and a much smaller spread in the transverse direction as shown in Fig.~\ref{fig:hadronproduction}.  Studies of the neutrino flux show a spectrum matching a stopped-pion decay-at-rest source with a minimal decay-in-flight component produced by the long graphite target.  Total $\pi^0$ production rates at 800 MeV were cross-checked against simplified beam target simulations using MCNP and found to agree within $\sim$~15\%.

\section{Reference Detector Design and Simulation}

\begin{figure}
\centering
  \includegraphics[width=0.6\textwidth]{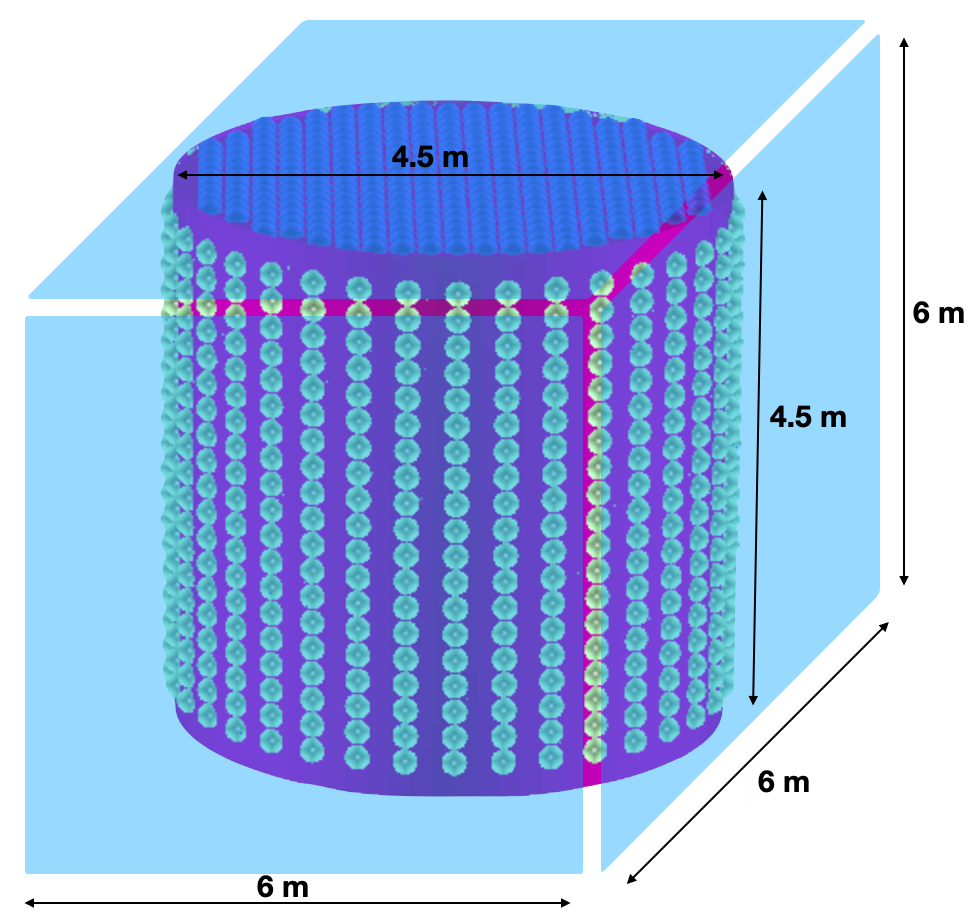}
  \caption{Rendering of the baseline liquid argon detector considered for the PIP2-BD experiment. The inner 4.5~m diameter, 4.5~m height cylindrical volume is enclosed within a 6x6x6~m$^3$ LAr cryostat. The fiducial mass of this detector is 100~tonnes.}
  \label{fig:setup}
\end{figure}

The reference PIP2-BD detector is a cylindrical, $\sim100$~tonne scintillation-only liquid argon (LAr) detector with a 5~m height and a 2.5~m radius positioned inside a larger LAr volume enclosed by a cubical membrane cryostat, which measures 6~m in each dimension and is instrumented as an active veto. The inner active volume is an optically isolated region defined by a reflective, 1/8'' thick teflon surface located 25~cm away from the edges of the detector. The teflon is coated with a layer of tetraphenyl butadiene (TPB) to shift VUV LAr scintillation photons into the visible range.  The scintillation light is detected by TPB-coated 8'' Hamamatsu R5912-MOD02 photomultiplier tubes (PMTs) located in holes cut out of the reflective teflon cylinder in both of the endcaps and the sidewalls.  There are 1294 PMTs total in the detector with 684 on the side wall of the teflon cylinder and 305 on each endcap. Fig~\ref{fig:setup} gives a rendering of the PIP2-BD inner and outer volumes.\\

A custom, Geant4-based (v4.10.6p01) simulation with a GDML description of the detector geometry was built to model the response of this detector. LAr scintillates in the VUV wavelengths with the peak emission at 128~nm with a scintillation yield of 40000~photons/MeV of energy deposited.  Within the simulation, these photons can populate two different emission states. The first is a singlet ("prompt") component with a lifetime of 6~ns and the second is a triplet ("late") component with a lifetime of 1600~ns.  The population of these states during the LAr scintillation is particle-dependent and selected via the definition and name of the particle generating the scintillation light.  For nuclear recoils, an energy-dependent ratio of prompt to late light is input that agrees with the values measured in Ref.~\cite{Regenfus2012} for nuclear recoils. In the electron recoil case an energy-independent prompt-to-late light ratio of 0.3 is input and this value is then adjusted in post-processing to an energy-dependent value where the central values agree with the values measured in Ref.~\cite{Regenfus2012} for electron recoils. The liquid argon quenching factor is modeled after the results from the COHERENT collaboration from the first measurement of CEvNS on liquid argon~\cite{COHERENT:2020iec}.\\

\begin{figure}
\centering
  \includegraphics[width=0.7\textwidth]{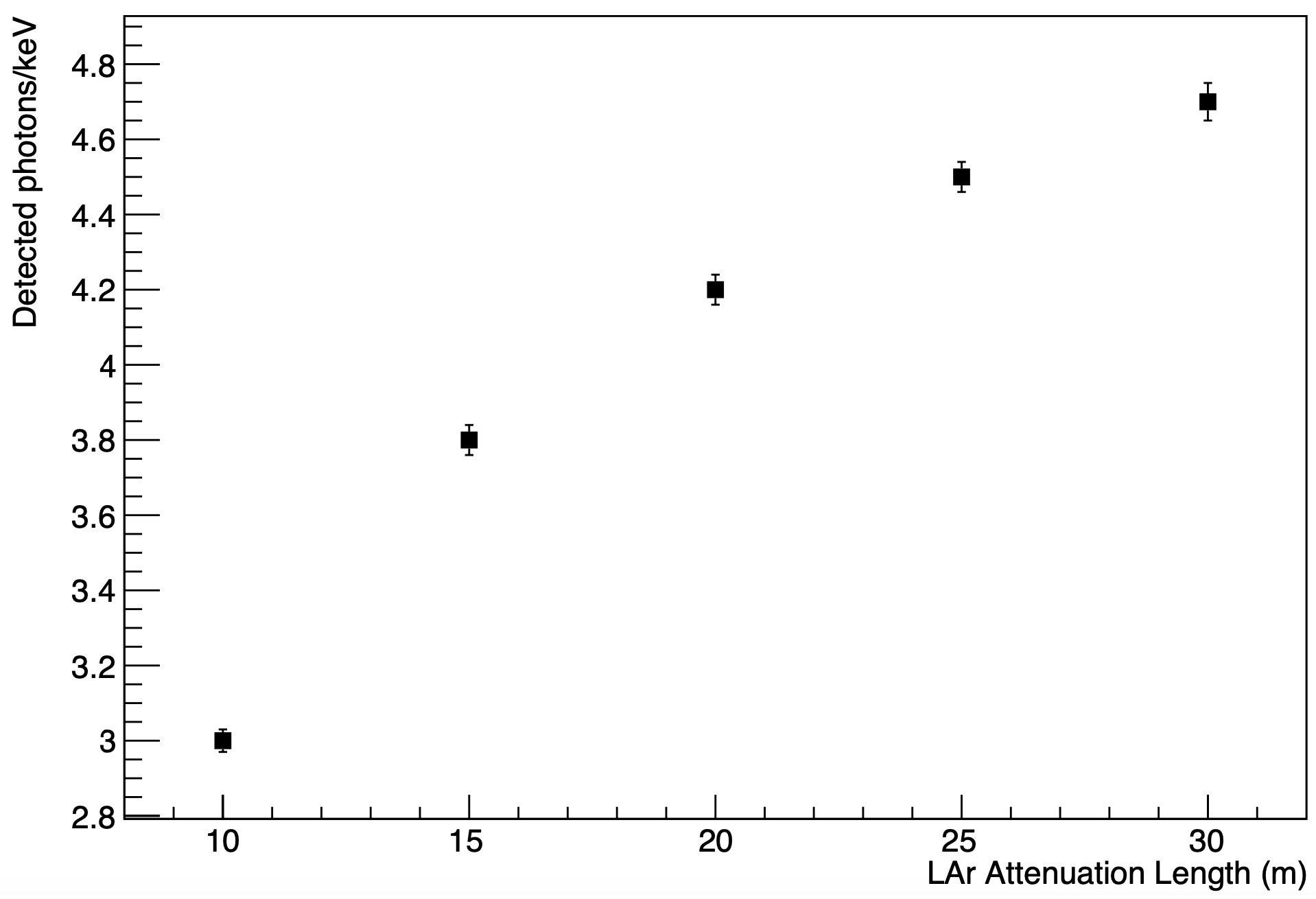}
  \caption{The resulting detected photon yield per keV of deposited energy for simulations with an adjusted liquid argon attenuation length. These simulations show the need for excellent liquid argon purity in order for the PIP2-BD detector to meet the goals of a 20~keVnr threshold. The resulting simulation assume a 20~m liquid argon attenuation length is achieved in the PIP2-BD detector.}
  \label{fig:lightyield}
\end{figure}

A 2~$\mu$s thick TPB layer is placed front of the reflective cylinder and around the active hemispherical face of the PMTs and wavelength-shifts the vast majority of the LAr scintillation that comes in contact with it.  The TPB emission spectrum within the simulation is taken from~\cite{Gehman:2011xm,Benson:2017vbw} and a 100\% conversion efficiency is assumed.  The teflon layer is set to have a 99\% reflectivity for visible wavelengths and to completely absorb any LAr scintillation light.  The PMTs are not sensitive to 128~nm light but have an $\sim18\%$ quantum efficiency at $\sim400$~nm wavelengths.  The 1 kHz PMT dark rate expected at LAr temperatures is also included in a post-processing stage and added to simulated events.  The LAr Rayleigh scattering length is wavelength-dependent and taken from Ref.~\cite{Grace:2015yta} with the values scaled such that the scattering length at 128~nm agrees with the measurement of 99.9~cm from Ref.~\cite{Gil-Botella:2021gco}. The refractive index of the LAr is also taken from Ref.~\cite{Grace:2015yta} and is also wavelength-dependent. The absorption length for all wavelengths is set to 20~m, representing the need for very high LAr purity as impurities decrease the absorption length from this nominal value. Fig.~\ref{fig:lightyield} shows the light yield for electrons simulated at the center of the detector as a function of absorption length to illustrate this effect.

\section{Signal and background generation}

\subsection{CEvNS}

We simulate CEvNS events throughout the detector for $\nu_\mu$, $\nu_e$, and $\bar\nu_\mu$ produced in $\pi^+ \rightarrow \mu^+ + \nu_\mu$ and subsequent $\mu^+ \rightarrow e^+ + \nu_e + \bar\nu_\mu$ decay.  The $\nu_\mu$ are simulated with an energy of $29.8$~MeV, while the energy spectra for the $\nu_e$ and $\bar\nu_\mu$ are sampled from the Michel spectra~\cite{Amanik:2009zz,Moreno:2015bta}
\begin{equation} \label{eqn:mudecayspect}
    \begin{split}
    f(E_{\nu_e}) & = \frac{96}{m_{\mu}^4}E_{\nu_e}^2(m_{\mu} - 2E_{\nu_e})dE_{\nu_e} \\
    f(E_{\bar\nu_\mu}) & = \frac{16}{m_{\mu}^4}E_{\bar\nu_\mu}^2(3m_{\mu} - 4E_{\bar\nu_\mu})dE_{\bar\nu_\mu}
    \end{split}
\end{equation}
with $0 \leq E_{\nu}\leq \frac{m_\mu}{2}=52.8$~MeV. Neutrino generation times are drawn according to the beam time structure and the 26~ns and 2.2~$\mu$s lifetimes of the $\pi^+$ and $\mu^+$, and the neutrino generation positions within the source are drawn according to a Gaussian distribution with a 30~cm width in each dimension.\\

\begin{figure}
\centering
  \includegraphics[width=0.7\textwidth]{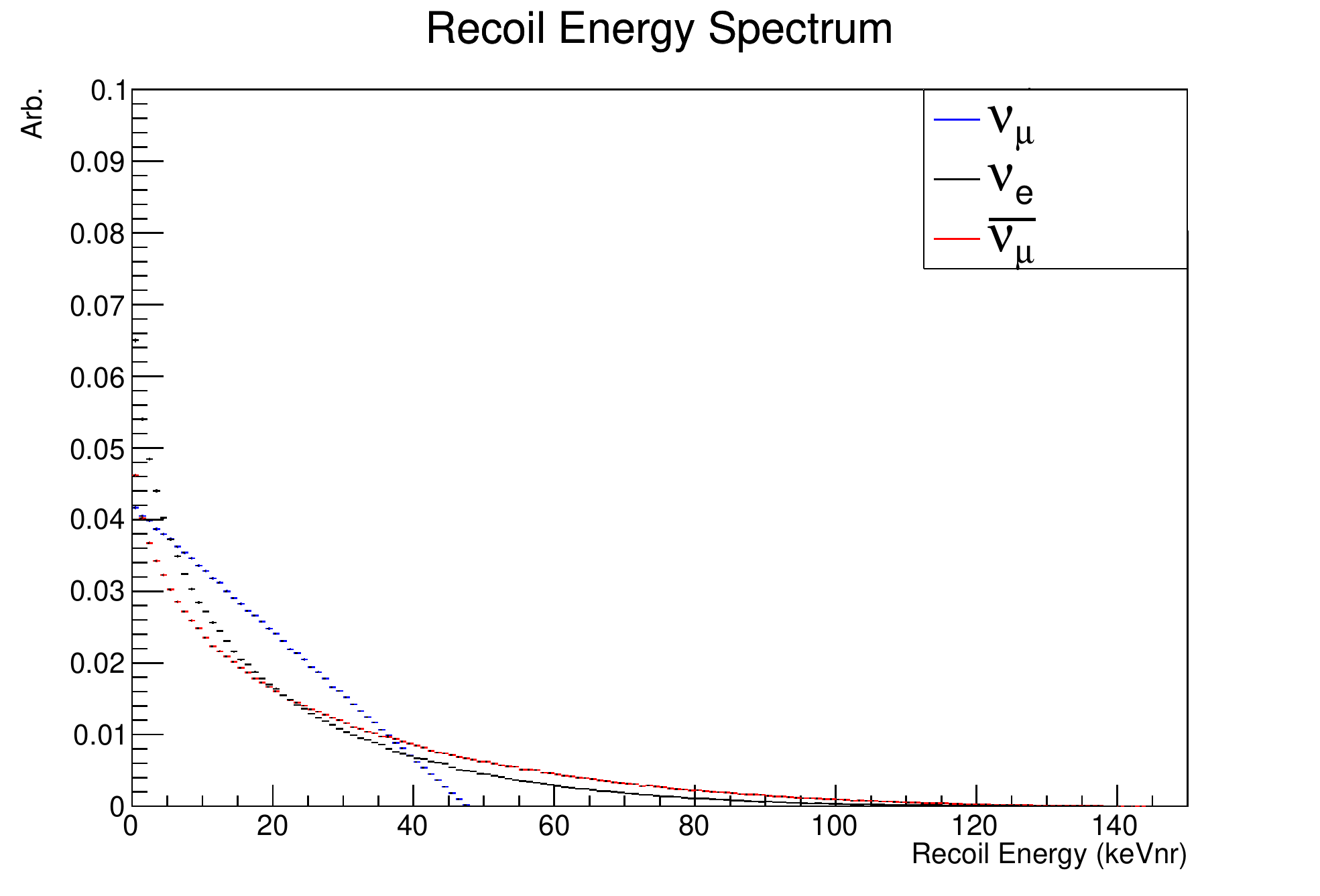}
  \caption{The recoil spectra for the three flavors of neutrinos predominantly produced with a stopped pion source. The $\nu_\mu$ is produced in the initial $\pi^+$ decay, while the $\nu_e$ and $\bar\nu_\mu$ are produced in the subsequent $\mu^+$ decay.}
  \label{fig:nuclearrecoils}
\end{figure}

Nuclear recoil energies for CEvNS events for each neutrino flavor are then sampled according to 

\begin{eqnarray}
\frac{d\sigma}{dT_{Ar}}&\sim&\frac{G_F^2M_{Ar}}{2\pi}\frac{Q_W^2}{4}(2-M_{Ar}T_{Ar}/E_\nu^2),
\label{eq:cevns}
\end{eqnarray}
where $Q_W=N-(1-4\sin^2\theta_W)Z$ and form factor suppression has been neglected. Fig.~\ref{fig:nuclearrecoils} shows the resulting distribution of nuclear recoils that are then used to re-weight a high-statistics sample of nuclear recoil events generated with a flat spectrum between 0--150~keV. An energy-dependent position resolution for CEvNS events is modeled as a Gaussian in each dimension with a width of $40\%/\sqrt{T(\textnormal{keV})/20}$ based on simulation studies of the performance of a simple reconstruction algorithm that computes the charge-weighted hit PMT position in the detector.

\subsection{$^{39}$Ar and PMT dark rate}

$^{39}$Ar is a cosmogenically-produced isotope present at the level of 1~Bq/kg in natural (atmospheric) argon. With an endpoint energy of 565~keV, $^{39}$Ar decays make up the main electron recoil background to nuclear recoil signals in a low threshold scintillation-only LAr detector. We use the Geant4 simulation to predict the level of $^{39}$Ar that contributes to the background. Two separate $^{39}$Ar simulations are performed in order to reduce the simulation time to manageable levels while providing a suitable sample size for our studies. The first is a small-scale simulation of 75k events of the full $^{39}$Ar beta-decay spectrum generated through the built-in decay macro commands in Geant4. The second simulates a flat spectrum of $^{39}$Ar beta-decays in the CEvNS region of interest of $\leq50$~keV. A much larger sample of events are generated through the second simulation and that allows for a similar computation time as for the first simulation. A reweighting of the first simulation to the second accounts for the fraction of the $^{39}$Ar spectrum considered in the energy range selected in the second simulation.\\

The $^{39}$Ar background cannot be shielded as it is produced throughout the active LAr volume.  And while levels of $^{39}$Ar have been measured in underground argon that are a factor of 1400 lower than in atmospheric argon, we assume here an $^{39}$Ar activity of 1~Bq/kg and use simple beam timing and pulse shape discrimination (PSD) to mitigate this background. PSD leverages the fact that nuclear and electron recoils populate the singlet and triplet LAr scintillation emission states in different fractions, so that the time distribution of the light in an event can be used to separate $^{39}$Ar electron recoil events from the nuclear recoil events.  When processing simulated events, the photon population is split into two populations based on the photon time $t_{\textnormal{hit}}$ with respect to the initial detection of light in the event. The ``prompt'' photons ($N_{\textnormal{prompt}}$) have $t_{\textnormal{hit}}\leq90$~ns. The ``delayed'' photons ($N_{\textnormal{delayed}}$) have $90<t_{\textnormal{hit}}<6000$~ns. The length of the delayed window takes into consideration the nominal lifetime of the liquid argon scintillation triplet state decay of 1600~ns to collect a large fraction of the light. The $F_{90}$ parameter, defined as
\begin{equation}
    F_{90} = \frac{N_{\textnormal{prompt}}}{N_{\textnormal{prompt}}+ N_{\textnormal{delayed}}},
\end{equation}
looks at the fraction of prompt light in an event. The $F_{90}$ parameter is energy-dependent but expected at $\sim0.3$ for electron recoil events such as $^{39}$Ar and $\sim0.7$ for nuclear recoil events such as CEvNS and DM. An energy dependent cut in the $F_{90}$ value follows the curve determined through a signal-to-background figure of merit analysis for the data analysis from the COHERENT Collaboration following Ref.~\cite{Zettlemoyer:2020kgh}. This curve is a reasonable proxy to study the effect of an $F_{90}$ cut on these simulations, although the central values of the $F_{90}$ distributions differ between the simulations from this study and the COHERENT data.\\

\begin{figure}
\centering
  \includegraphics[width=0.48\textwidth]{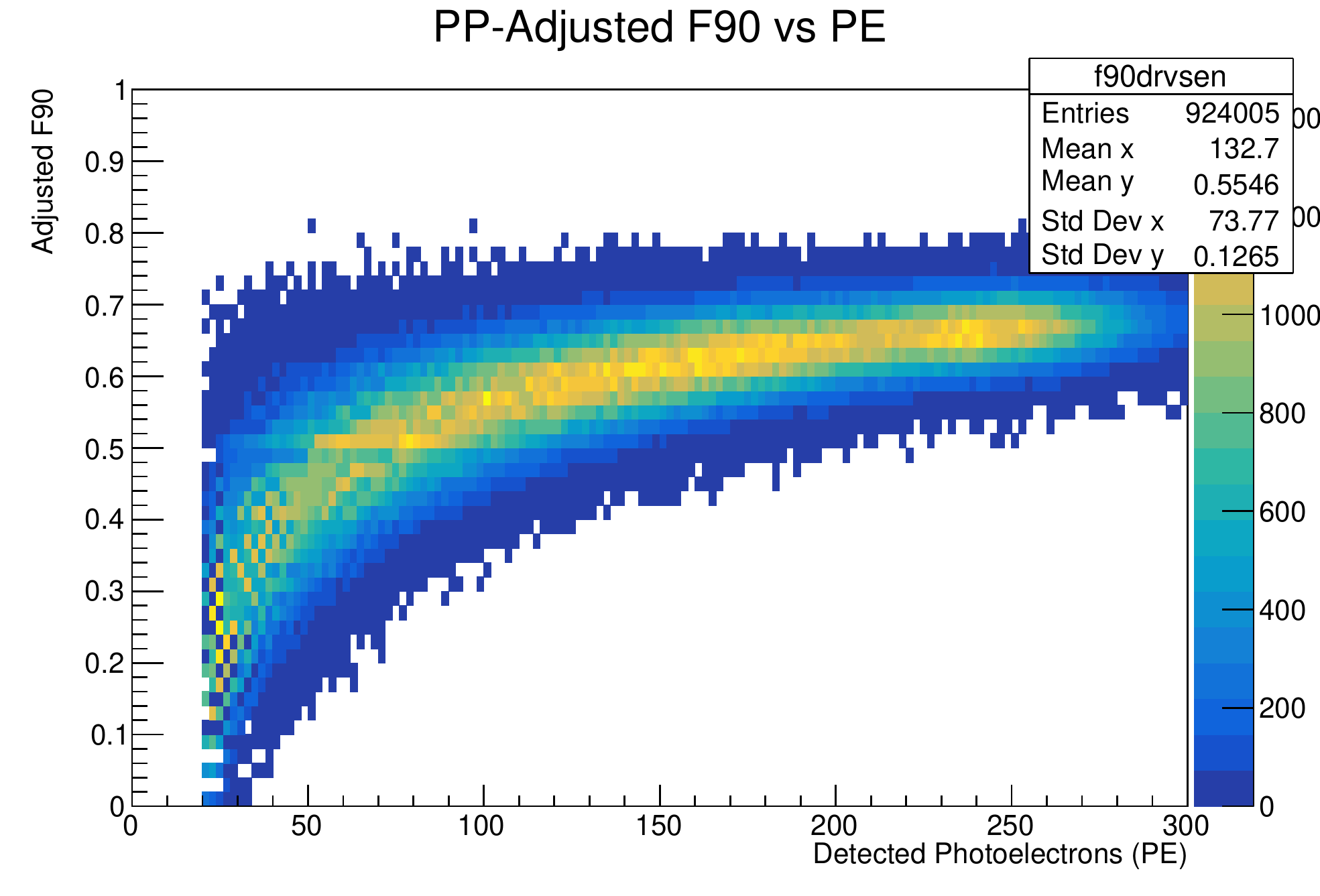}
  \includegraphics[width=0.48\textwidth]{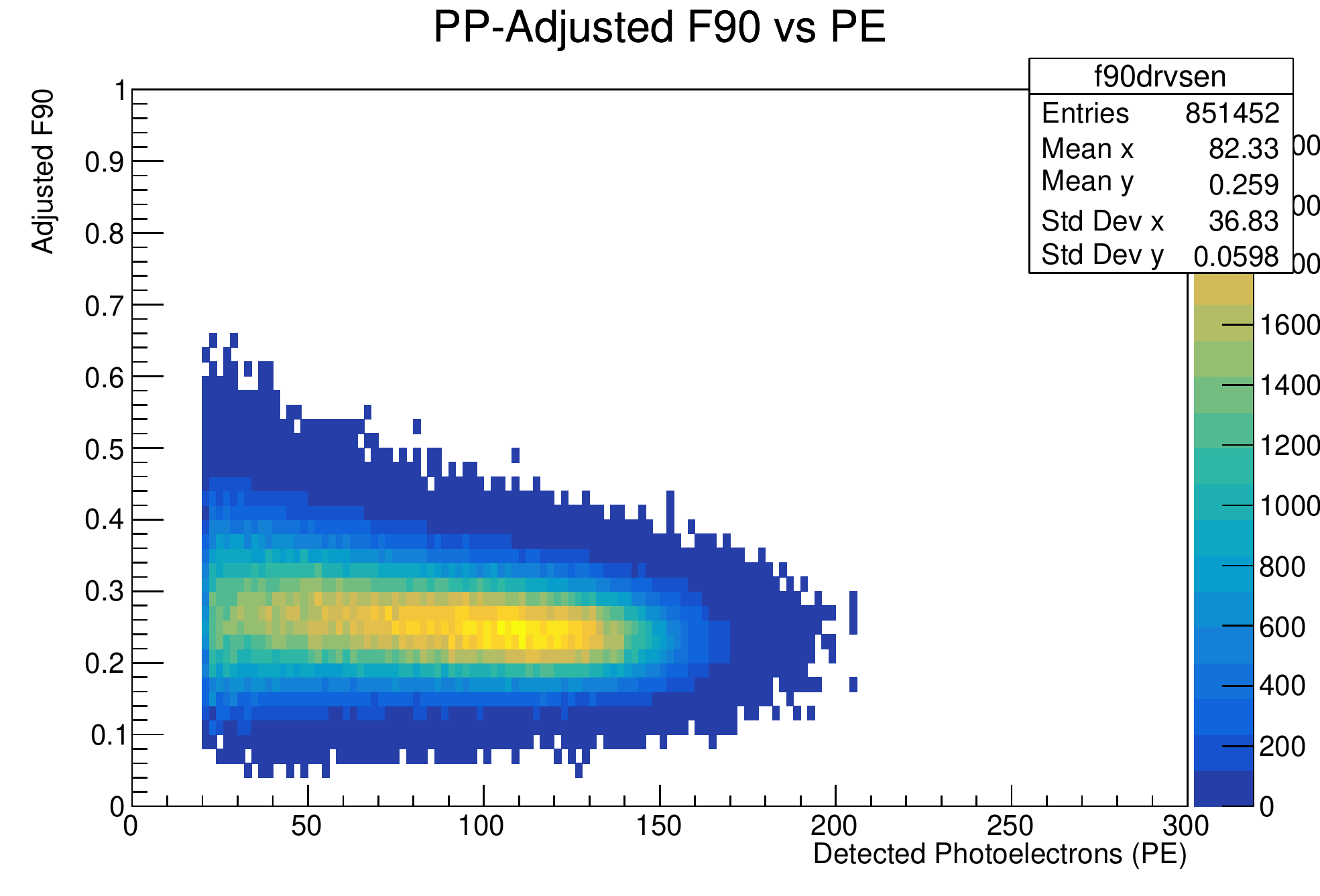}
  \caption{Resulting $F_{90}$ distributions for simulated CEvNS events (left) and $^{39}$Ar (right). $F_{90}$ is the fraction of photons detected within the first 90~ns of the events and used as a pulse shape discrimination parameter in liquid argon.}
  \label{fig:F90}
\end{figure}

PMT dark counts can also contribute background in our detector, given the large number of 8'' PMTs and the 6~$\mu$s delayed window. We find that this background is easily mitigated using a simple cut that removes any event with $N_{\textnormal{PE}}<20$, which corresponds to a reconstructed nuclear recoil energy of $\sim20$~keV. PMT dark rate, assumed here as 1~kHz, also produces a noticeable shift in the $F_{90}$ distributions of electron and nuclear recoil events to lower values.  Therefore, we incorporate a constant shift of 0.08 in the cut towards lower values of $F_{90}$ values to account for the effects of dark count rate. Fig.~\ref{fig:F90} shows the $F_{90}$ distributions for CEvNS and $^{39}$Ar events after the inclusion of PMT dark count rate and removing events with $N_{\textnormal{PE}}<20$.\\

\subsection{Vector Portal Dark Matter}

Ref.~\cite{deNiverville:2011it,deNiverville:2015mwa,deNiverville:2016rqh} describes the vector-portal DM signal that forms the basis of the signal prediction. The detection signature for the DM particles are also nuclear recoils of the DM particle $\chi$ with the nucleus. Therefore, the DM signal is also a low-energy nuclear recoil which makes CEvNS a background for such a DM search. The parameter of interest for our studies is defined as the dimensionless variable
\begin{equation}
    Y = \epsilon^2\alpha_D(\frac{m_\chi}{m_V})^4, 
\end{equation}
where $\epsilon$ parameterizes the kinetic mixing between the SM photon and a dark, heavy photon, $\alpha_D$ is the dark fine structure constant (set to a value of 0.5 in this study), and the mass of the heavy photon, $m_V$, is fixed to the value of 3$m_\chi$.\\

\begin{figure}
\centering
  \includegraphics[width=0.9\textwidth]{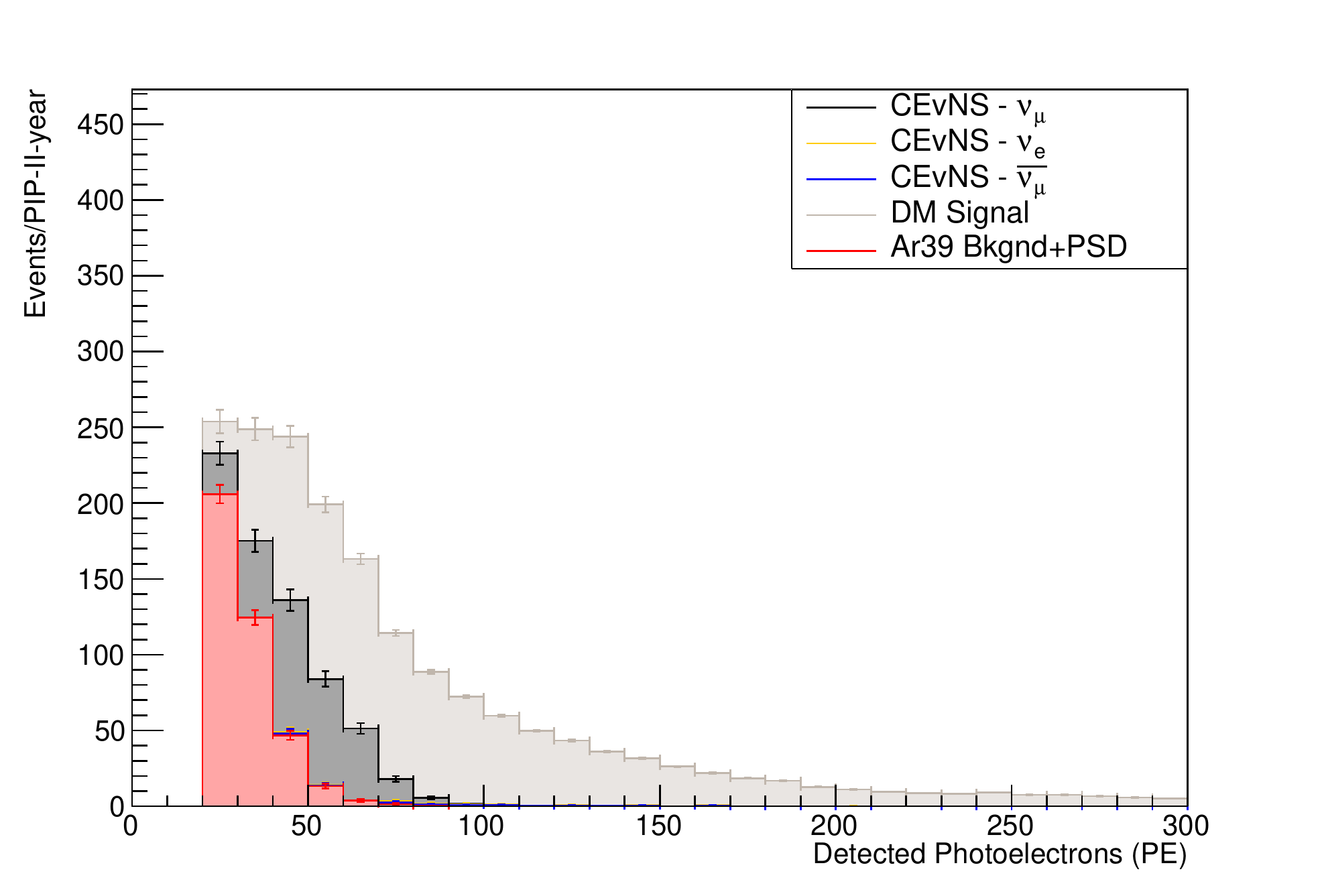}
  \caption{Stacked histogram showing the predicted event rates per year of PIP2-BD operation in the C-PAR scenario. The 20~ns pulse length envisioned in this scenario allows for excellent background rejection of the CEvNS neutrino and idealized $^{39}$Ar background components while preserving the ultra-prompt DM signal.}
  \label{fig:dmsignal}
\end{figure}

The BdNMC simulation code, described in detail in Ref.~\cite{deNiverville:2016rqh}, generates the initial DM nuclear recoil predictions. BdNMC takes as an input the locations, energies, and momenta of $\pi^0$ produced by the proton collisions with the target from the beam simulation as well as the dimensions and positioning of the detector in the coordinate system defined by the beam simulation, which for PIP2-BD is 18 m downstream of the beam target, on-axis. The nuclear recoil events output by BdNMC are then fed into the full detector simulation. A set of 375,000 events was generated for $m_\chi=[1,5,10,30,50]$~MeV at fixed values of $\epsilon=4.04\times10^{-4}$, $\alpha_D=0.5$, and $m_V=3m_\chi$, which can then be reweighted to other values of $\epsilon$.  Fig~\ref{fig:dmsignal} shows a histogram of DM nuclear recoil signal events for the C-PAR scenario stacked on top of CEvNS and $^{39}$Ar background events for a benchmark point of $m_\chi=10$~MeV and $\epsilon = 5.82 \times 10^{-5}$, corresponding to $Y=2.09 \times 10^{-11}$.\\

\section{Physics Reach}

\subsection{Dark Matter}

Recent theoretical work has identified several classes of sub-GeV DM, which are neutral under the Standard Model (SM) but charged under new ``dark sector'' forces that mediate DM interactions with SM particles and which achieve the correct relic abundance in the early universe through a standard thermal-freeze out mechanism~\cite{Battaglieri:2017aum}. What is most exciting about these classes of DM models is that, in the case that the mediator is heavier than the DM and the DM accounts for the thermal relic abundance, they predict a SM-DM coupling as a function of DM mass that is tantalizingly close to current limits and can be probed by next-generation accelerator-based fixed target dark sector searches in the coming decade~\cite{Batell:2009di,deNiverville:2011it,deNiverville:2012ij,Gardner:2015wea,Berlin:2018pwi,Tsai:2019buq,Berlin:2018bsc,Batell:2018fqo,NA64:2017vtt,FASER:2019aik,Alekhin:2015byh,NA62:2017rwk}. Proton beam dump experiments are potentially sensitive to these and any other dark sector models that produce light DM directly through hadronic interactions or through the subsequent decay of light mesons. This includes, for example, both standard ``vector portal'' DM models that can be probed with both proton and electron beams as well as other models, such as hadrophilic DM, for which proton beams provide unique sensitivity~\cite{Batell:2018fqo}.  Recent results from the COHERENT~\cite{COHERENT:2020iec} and Coherent CAPTAIN-Mills (CCM) experiments~\cite{CCM:2021leg,Aguilar-Arevalo:2021sbh} have demonstrated how detectors capable of measuring coherent elastic neutrino-nucleus scattering (CEvNS) can also be used to set limits on vector portal and leptophobic DM at proton beam dumps.

\subsubsection{Vector portal kinetic mixing}

We compute our rate-only sensitivity to the minimal vector portal kinetic mixing model at each DM mass point $m_\chi$ using a simple $\Delta\chi^2$ test statistic of the form:
\begin{equation} \label{eqn:chi2}
    \Delta\chi^2 = \frac{N_{\textnormal{sig}}^2}{N_{\textnormal{bkg}} + \sigma^2N_{\textnormal{CEvNS}}^2},
\end{equation}
where we have assumed that the $^{39}$Ar background rate can be measured outside of beam windows to a very high precision so that only a systematic uncertainty on the normalization of beam-related CEvNS events has been added to the usual statistics-only $\Delta\chi^2$ definition. However, because the DM signal is very prompt, beam timing can also be used to measure and constrain the CEvNS rate from delayed neutrinos outside of the beam spill window. Therefore, we use the statistical uncertainty on the CEvNS rate measured outside the beam spill to estimate the systematic uncertainty of the CEvNS rate during the beam spill. To obtain the 90\% C.L. for the dimensionless variable $Y$ at each DM mass point, we weight our generated events by $\epsilon^4$ until we find the value that gives $N_{\textnormal{sig}}$ total events such that $\Delta\chi^2=2.71$.\\

\begin{figure}
\centering
  \includegraphics[width=0.55\textwidth]{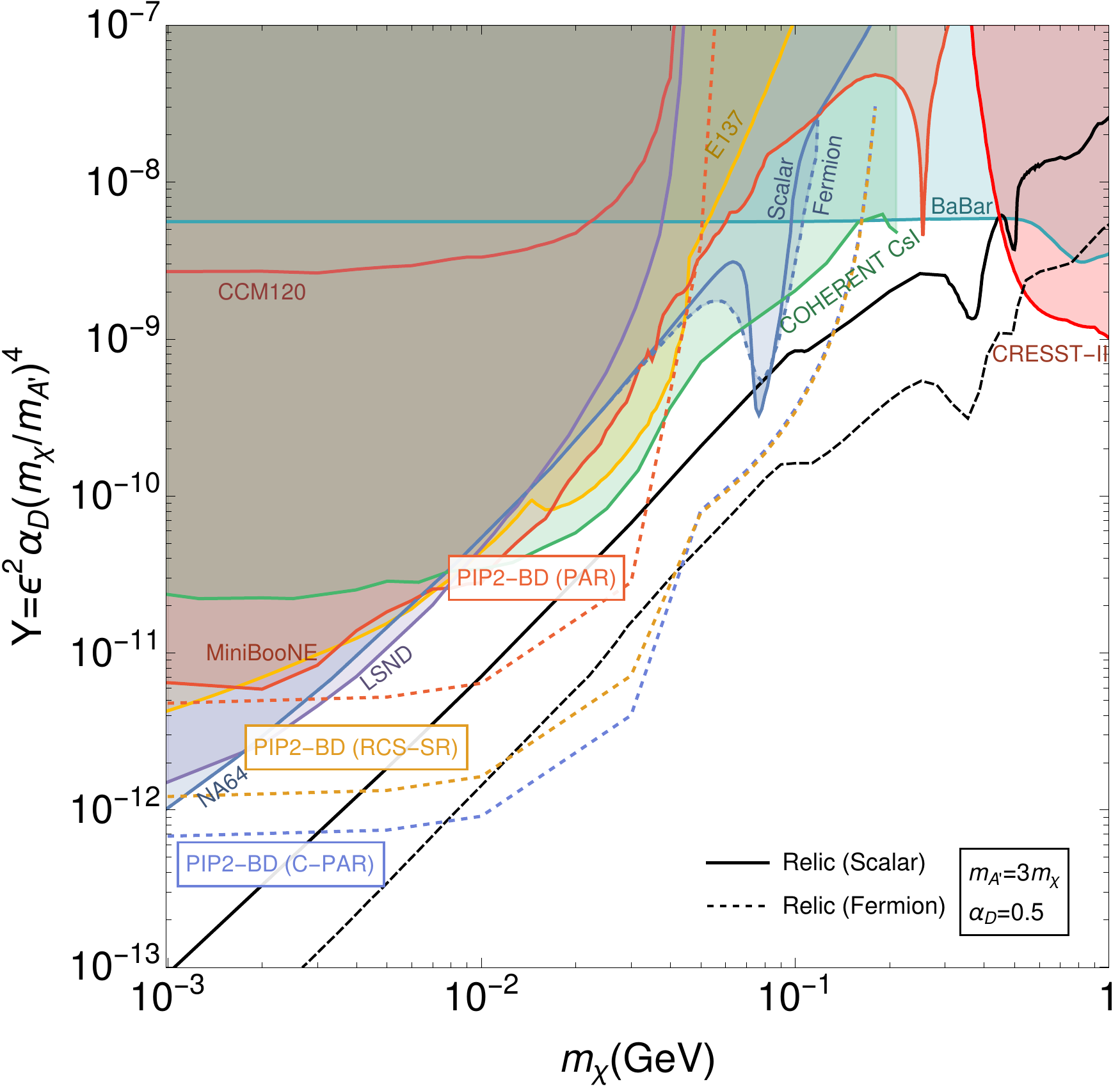}
  \caption{The 90\% C.L. sensitivity to the vector portal DM model for the three accumulator ring scenarios. Both RCS-SR and C-PAR are able to search a large parameter space reaching the expected thermal relic density for both scalar and fermion DM.}
  \label{fig:darkmatter}
\end{figure}

For the C-PAR and RCS-SR scenarios, the calculation was extended to heavier DM masses by also tracking eta production in the beam simulation and providing that to the BdNMC DM generator. After checking that the signal efficiency stayed relatively stable for a few DM mass points above $m_\chi=50$~MeV, a constant signal efficiency of 75\% was assumed for all DM mass points above $50$~MeV. This allowed the predicted number of events from the BdNMC to be used directly in the $\Delta\chi^2$ computation above, rather than first running it through the detector simulation. Fig.~\ref{fig:darkmatter} shows the 90\% C.L. sensitivities we obtain for all 3 accumulator ring scenarios for 5 years of running, compared to existing limits. We probe relic density targets for both scalar and fermion DM.

\subsubsection{Leptophobic model}

Leptophobic DM is an example of a class of dark sector models that couples preferentially to quarks rather than leptons, leading to a phenomenology that is similar but complementary to the above scenario and is subject to a different set of existing experimental limits.\\

In this model, baryon number is gauged rather than introducing a new $U(1)'$ gauge symmetry giving the following terms in the Lagrangian
\begin{equation}
    \mathcal{L}_B \supset - A^\mu_B (g_B J^B_\mu + g_\chi J^\chi_\mu + \varepsilon_B e J_\mu^\mathrm{EM}),
\end{equation}
where $J_\mu^B \equiv \frac{1}{3} \sum_i \bar q_i \gamma_\mu q_i$ is the SM baryonic current, $J_\mu^\chi$ is the dark sector current, and $J_\mu^\mathrm{EM}$ is the SM electromagnetic current. A thermal relic density target can also be defined for this model assuming $\alpha_B=eg_B/(4\pi)^2$.\\

\begin{figure}
\centering
  \includegraphics[width=0.75\textwidth]{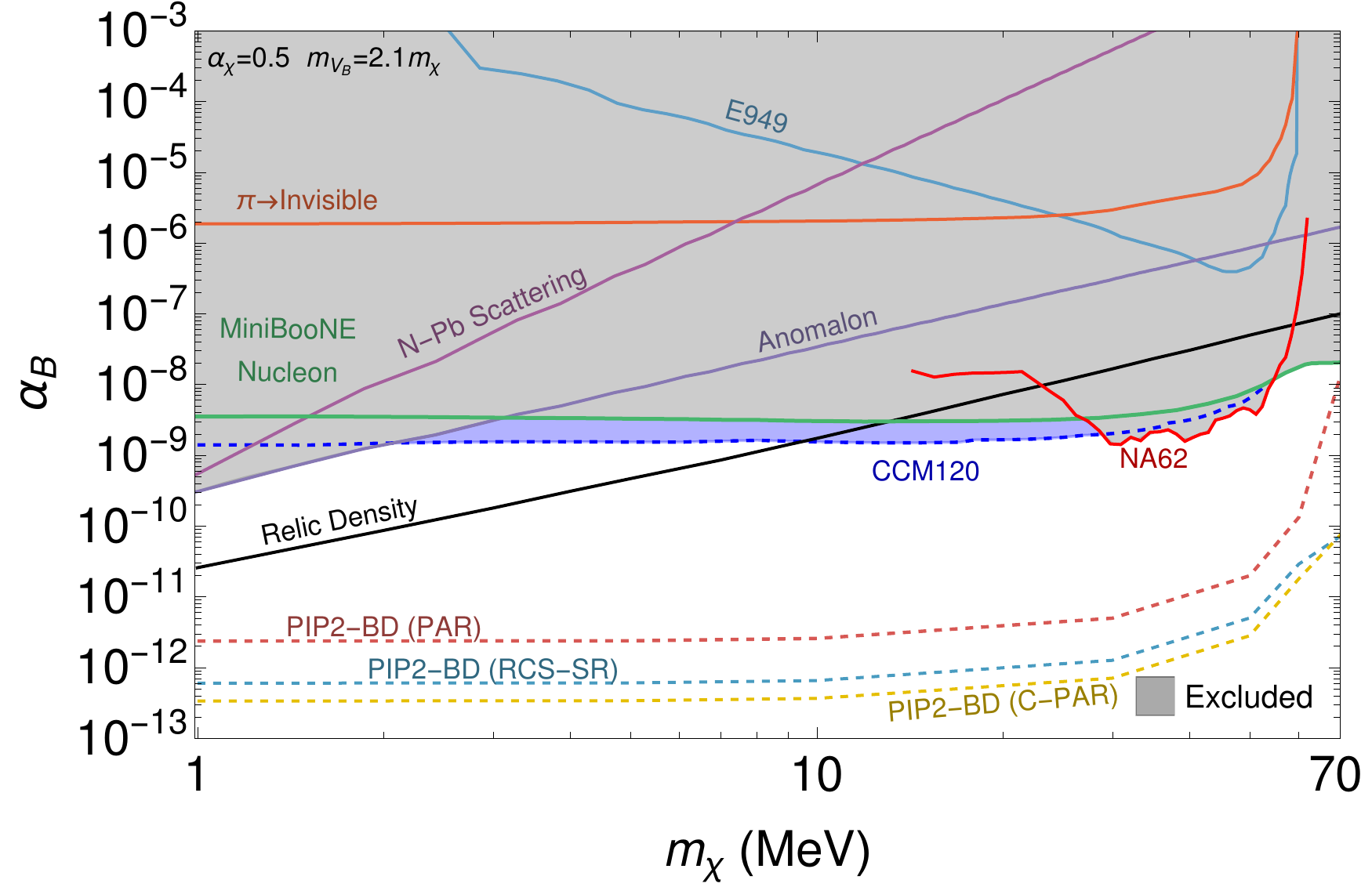}
  \caption{The 90\% C.L. sensitivity to the leptophobic DM models for each of the three accumulator ring scenarios.}
  \label{fig:leptophobic}
\end{figure}

This model predicts the same DM nuclear recoil energy distributions as the minimal vector portal model but with a rate that scales as $\alpha_\chi\alpha_B^2$ rather than $\alpha_D\epsilon^4$. We can therefore use signal rate predictions for leptophobic DM scattering from BdNMC to re-interpret the minimal vector portal sensitivities computed above in terms of $\alpha_B$ rather than $Y$. Fig.~\ref{fig:leptophobic} shows the 90\% C.L. sensitivities we obtain for all 3 accumulator ring scenarios for 5 years of running, compared to existing limits.

\subsubsection{Inelastic dark matter}

Interesting new phenomenology appears if the minimal vector portal scenario is extended to include two DM particles, $\chi_1$ and $\chi_2$ with $\Delta=(m_{\chi_2} - m_{\chi_1})/m_{\chi_1}>0$. Depending on the parameters of the model, $\chi_2$ may be sufficiently long-lived to propagate to the detector and then be observed through $\chi_2 \rightarrow \chi_1 e^+ e^-$ decay. If this process is not kinematically allowed, the DM may instead be observed through its up- or down-scattering off of electrons in the detector.\\

\begin{figure}
\centering
  \includegraphics[width=0.95\textwidth]{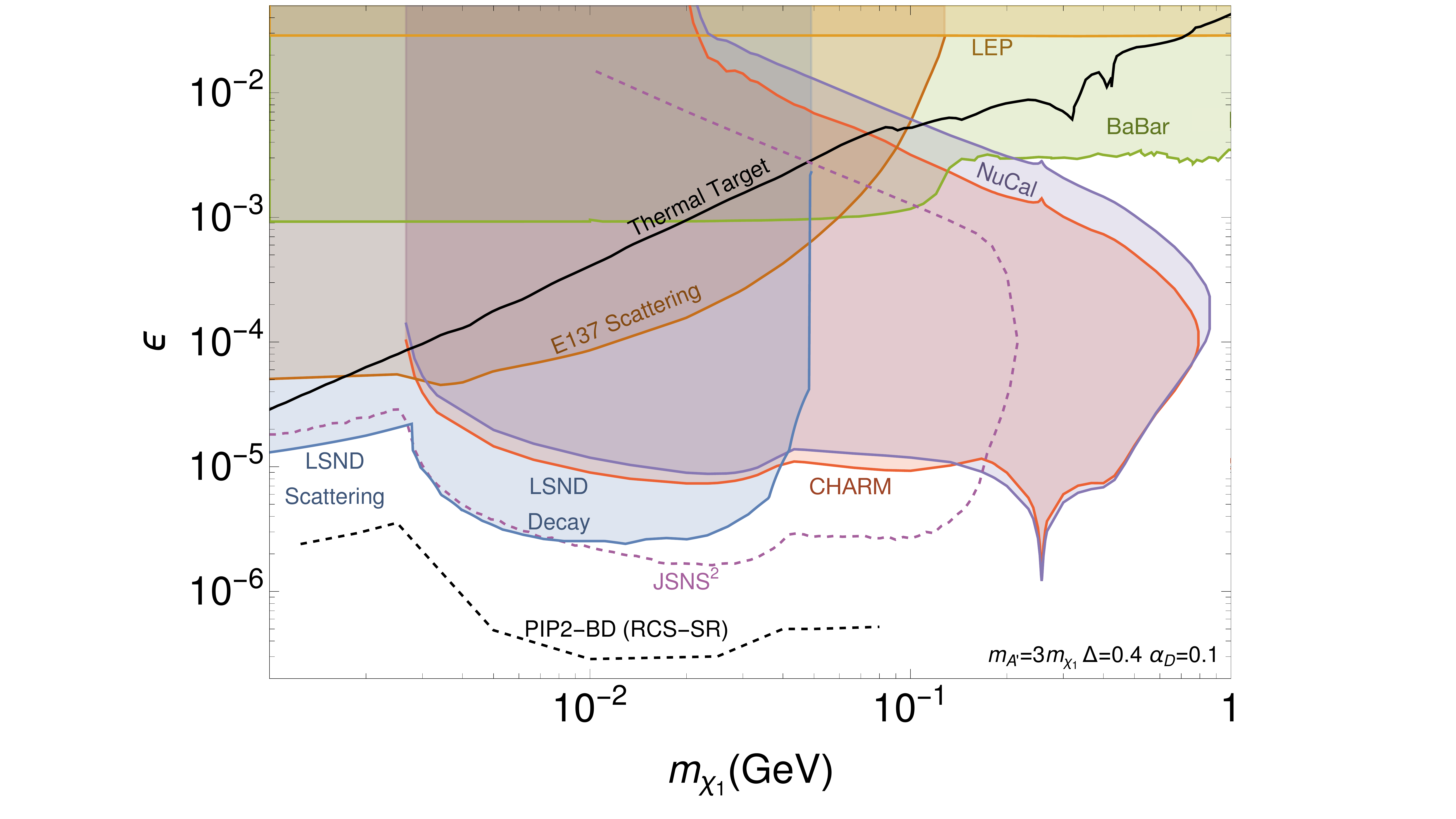}
  \caption{The three-event sensitivity for the RCS-SR accumulator ring scenario for inelastic DM. This sensitivity estimate does not include background predictions as the other DM sensitivity limits.}
  \label{fig:iDM}
\end{figure}

A recent study of this model in the JSNS$^2$ detector~\cite{Jordan:2018gcd} demonstrated how this signal can be isolated from background in a scintillation-only detector. Here, we make no attempt to quantify the expected backgrounds and simply use BdNMC to predict signal rates for both decay and scattering signatures.  The 3 event sensitivity curve obtained from BdNMC for 5 years of data-taking with the RCS-SR scenario is shown in Fig.~\ref{fig:iDM} along with existing constraints and the JSNS$^2$ sensitivity estimate.

\subsection{Axion-like particles}

Proton beam dumps produce not only a large quantity of hadrons such as pions, but also a large quantity of other particles, including photons, electrons, and positrons. Therefore, they also provide excellent opportunities to search for axion-like particles (ALPs)~\cite{Dent:2019ueq,AristizabalSierra:2020rom,Brdar:2020dpr}. We consider here ALP interactions parameterized by adding the following terms to the SM Lagrangian:
\begin{equation}
\mathcal{L}_{ALP} \supset -\frac{g_{a\gamma}}{4} a F_{\mu\nu}\tilde{F}^{\mu\nu}-ig_{ae}a \bar{e}\gamma^5 e.
\end{equation}

\begin{figure}
\centering
\includegraphics[width=0.5\textwidth]{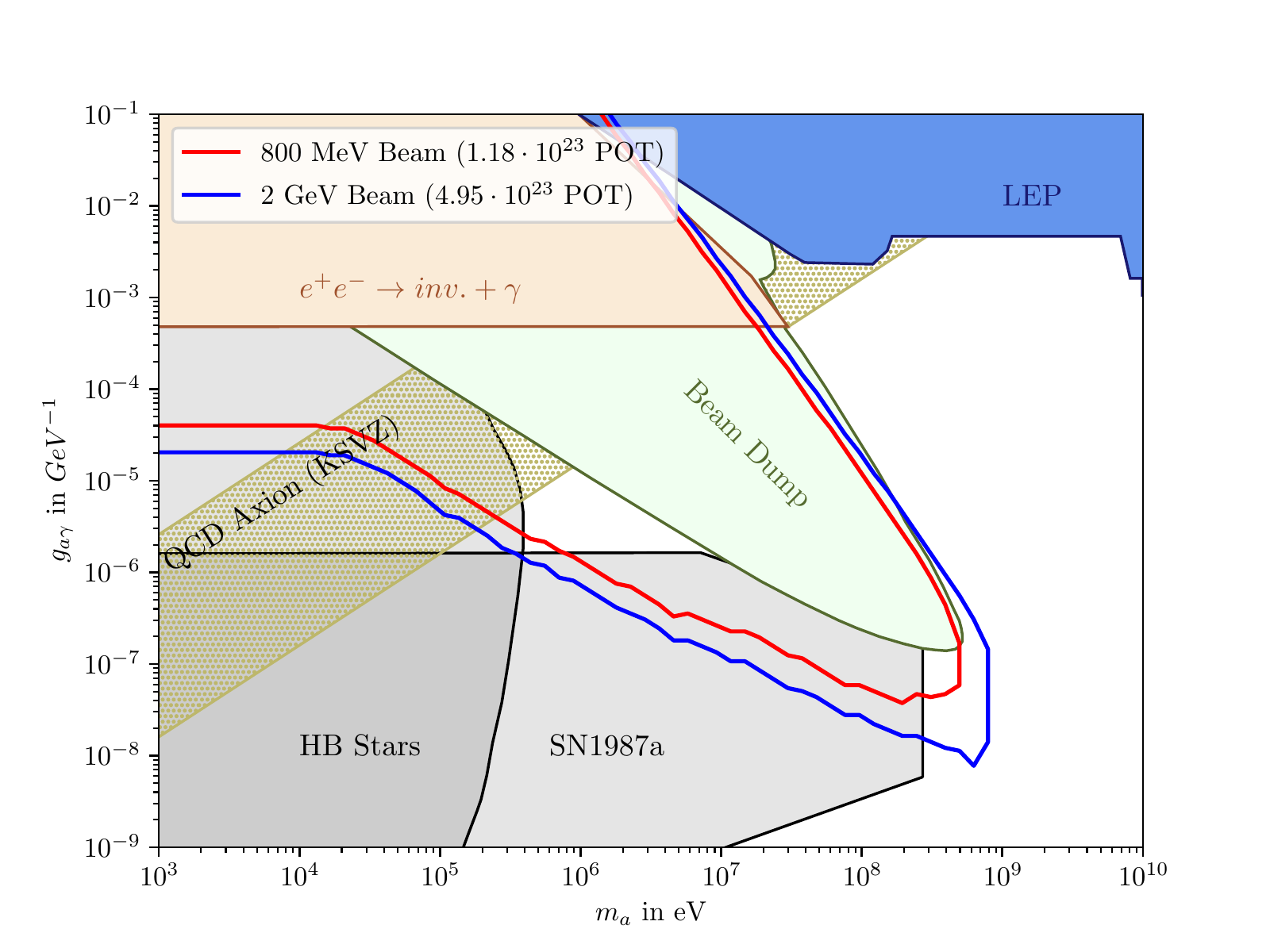}
  \includegraphics[width=0.46\textwidth]{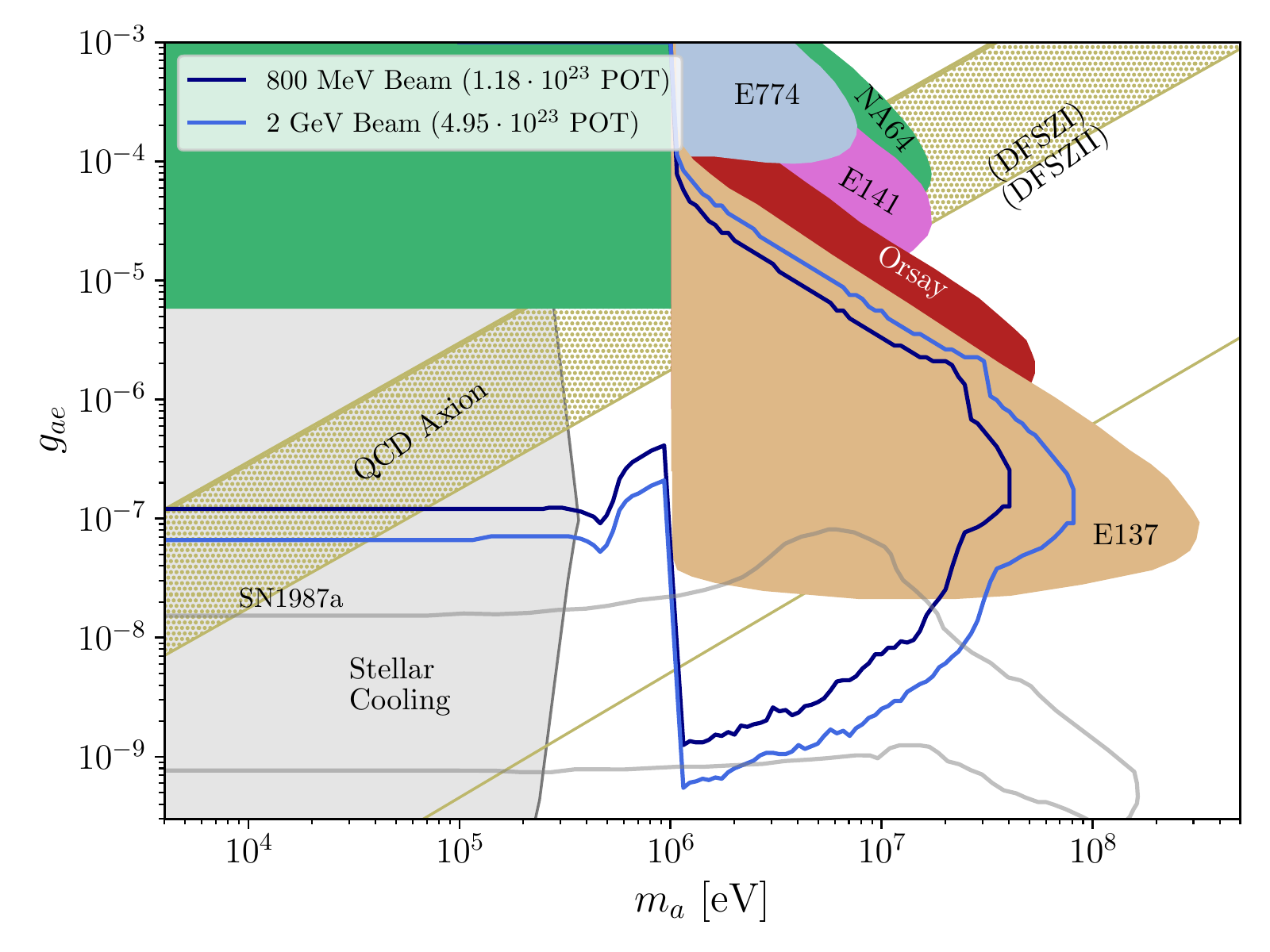}    \caption{Statistics-only 90\% C.L. sensitivity estimates for the axion-like particle (ALP) model. In five years of running, PIP2-BD can scan the $g_{a\gamma}$ ``cosmological triangle" region without existing constraints as well as a large part of a similarly untested parameter region for $g_{a e}$. These sensitivity estimates do not include background predictions.}
  \label{fig:ALP}
\end{figure}

If $g_{a\gamma}$ is non-zero, ALPs can be produced in the beam dump via the Primakoff processes (with a factor of $Z^2$ enhancement), propagate to the detector, and then either decay into two photons or produce a single photon via inverse Primakoff scattering.  On the other hand, if $g_{a e}$ is non-zero, ALPs can be produced in the beam dump via Compton-like scattering, bremsstrahlung, or $e^+e^-$ annihilation, propagate to the detector, and then decay to $e^+e^-$ or produce an electron and photon via inverse Compton-like scattering.

To compute the expected signal rate in PIP2-BD for the PAR and RCS-SR scenarios, we first obtain the photon, electron, and positron fluxes above 100 keV from the proton beam dump simulation. We then convolve this with the production cross-sections for the processes mentioned above and calculate the probability that the ALP decays or interacts in the detector, following Ref.~\cite{Brdar:2020dpr}. We scan over values of $m_a$ and determine the three event sensitivity curves for $g_{a\gamma}$ and $g_{a e}$, considering only one coupling nonzero at a time.  This corresponds to 2.3 expected signal events for 5 years of data-taking, assuming a 75\% efficiency for events above 100 keV, which is in line with the analysis used for the CCM scintillation-only LAr detector~\cite{CCM:2021lhc}.  The corresponding background-free, statistics-only 90\% C.L. ALP sensitivity limits are shown in Fig.~\ref{fig:ALP}.  In 5 years of data-taking, PIP2-BD will cover the $g_{a\gamma}$ ``cosmological triangle'' region as well as a large part of a similar parameter region for $g_{a e}$ where no lab-based constraints exist to rule out axion-like particles.

\subsection{Neutrinos}

\subsubsection{Testing short-baseline anomalies}

While both DM and ALPs have recently been invoked to explain the excess events observed at short baselines in both LSND and MiniBooNE, CEvNS provides a unique tool to also study the sterile neutrino oscillation hypothesis.  CEvNS can definitively establish the existence of sterile neutrinos through active-to-sterile neutrino oscillations~\cite{Anderson:2012pn}.  Using CEvNS, we can explore both mono-energetic $\nu_\mu$ disappearance with $\text{E}_\nu$ = 30~MeV and the summed disappearance of $\nu_\mu$, $\bar\nu_\mu$, and $\nu_e$ to $\nu_S$, which can also put constraints on $\nu_\mu \rightarrow \nu_e$ oscillation parameters in a 3+1 sterile neutrino model.\\

\begin{figure}
\centering
  \includegraphics[width=0.32\textwidth]{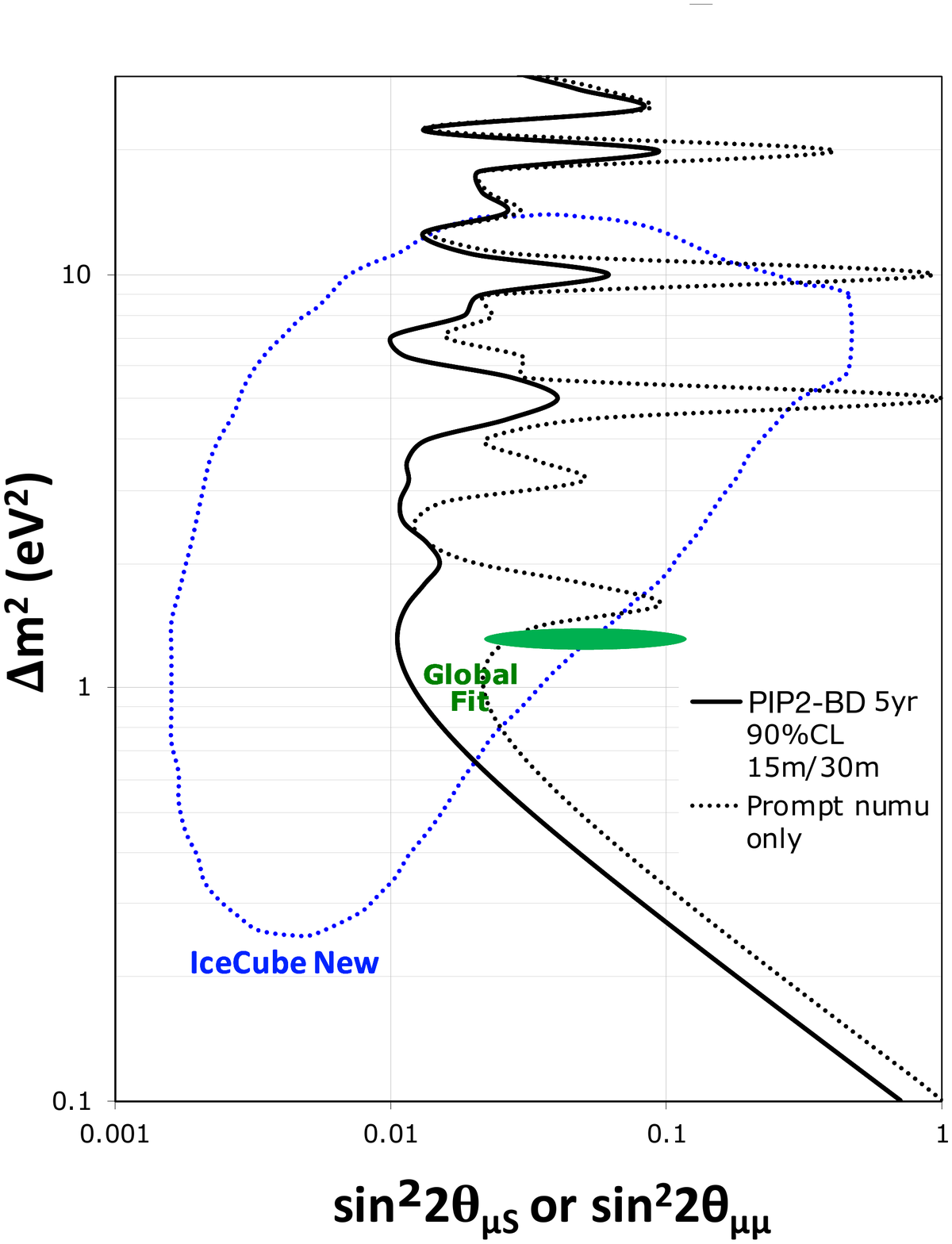}  \includegraphics[width=0.32\textwidth]{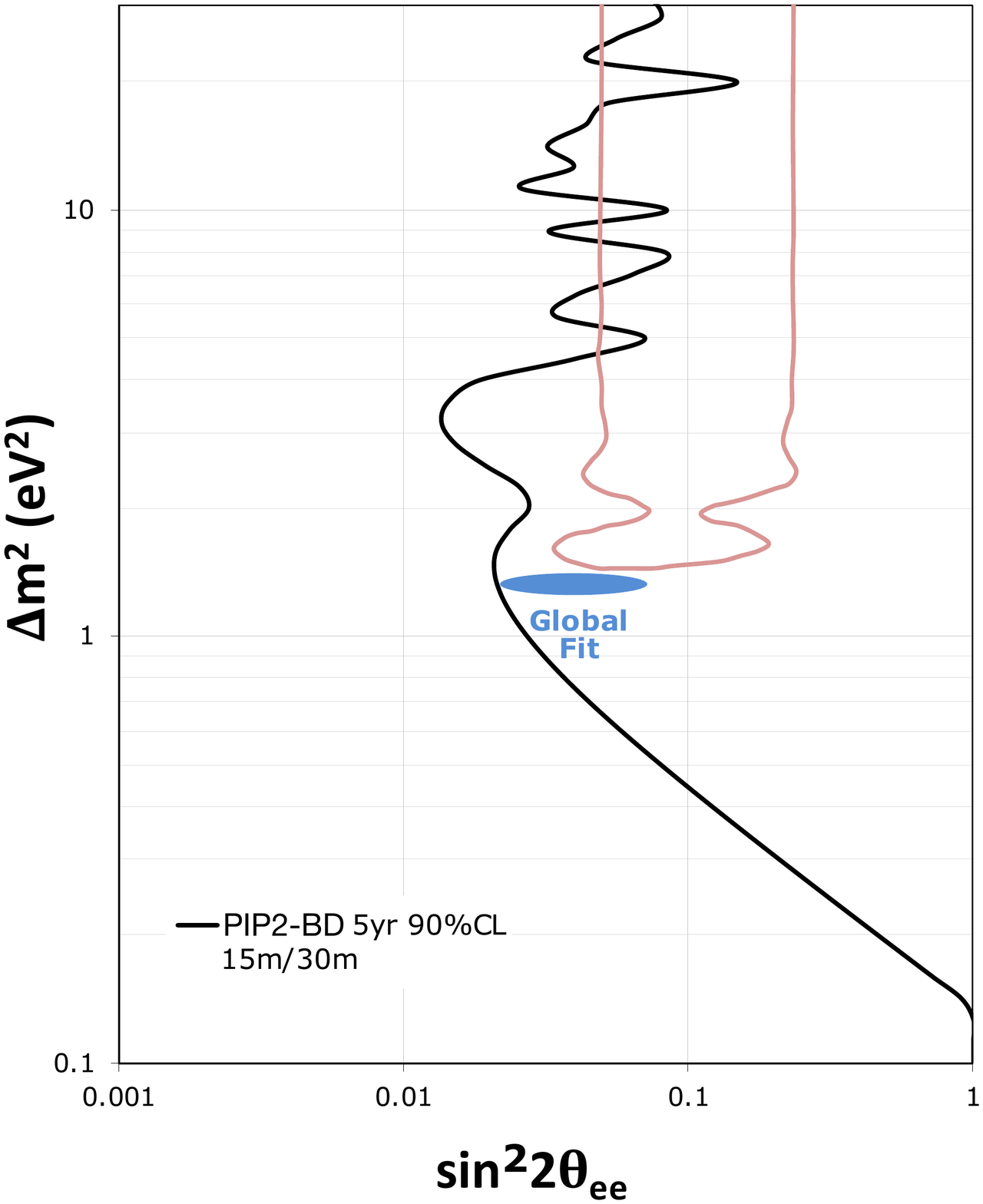}  \includegraphics[width=0.32\textwidth]{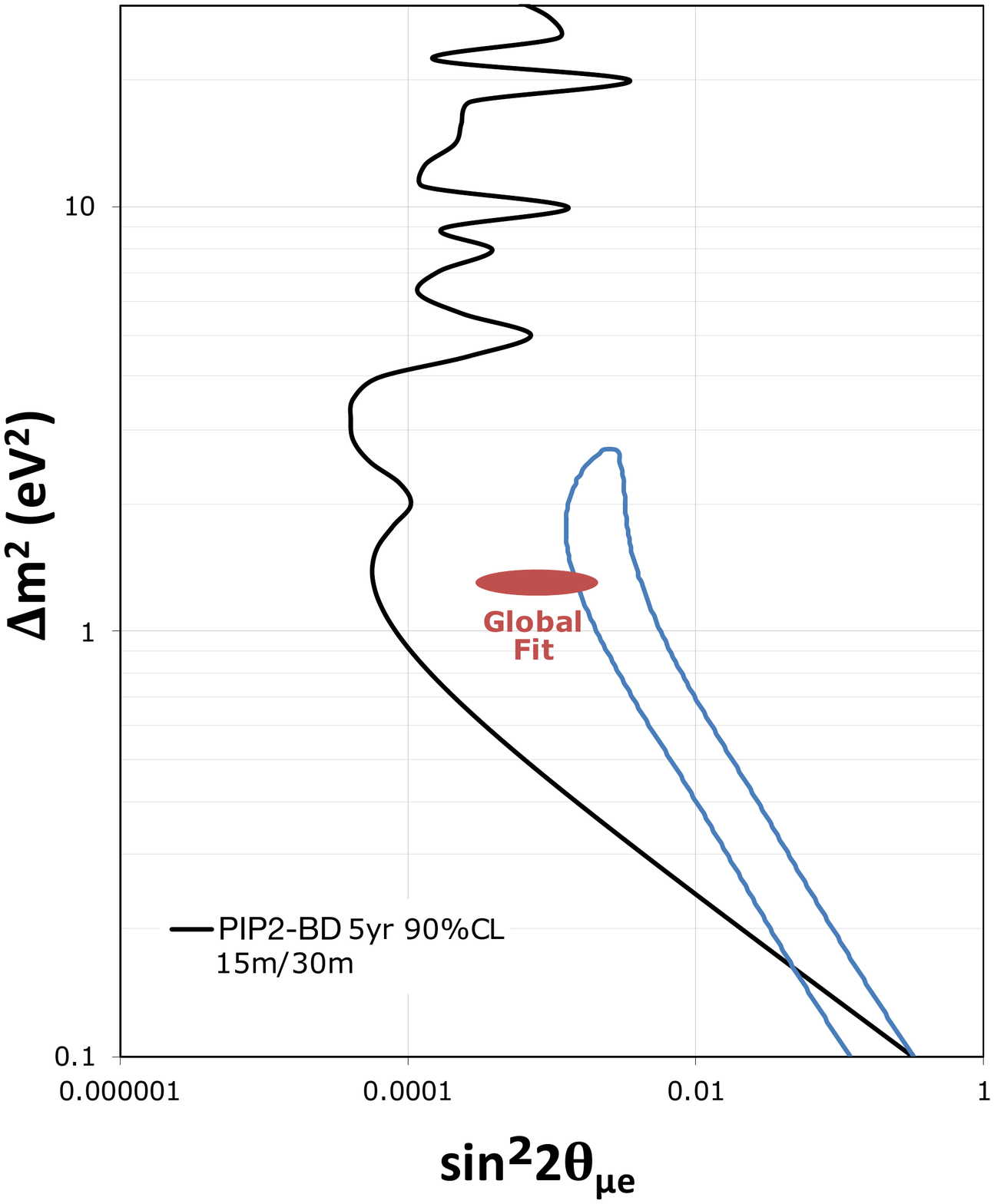}
  \caption{PIP2-BD 90\% confidence limits on active-to-sterile neutrino mixing compared to existing $\nu_\mu$ disappearance limits from IceCube~\cite{Aartsen:2020iky} and a recent global fit~\cite{Diaz:2019fwt}, assuming a 5~year run (left). Also shown are the 90\% confidence limits for $\nu_\mu$ disappearance (left), $\nu_e$ disappearance (middle), and $\nu_e$ appearance (right), assuming the $\bar\nu_\mu$ and $\nu_e$ can be detected with similar assumptions as for the $\nu_\mu$. }
  \label{fig:sterilenu}
\end{figure}

We consider here a PIP2-BD setup consisting of two identical detectors, located 15~m and 30~m away from the beam dump with a 20~keV recoil energy threshold and an efficiency of 70\%. We assume the neutron background in this dedicated facility could be suppressed to a negligible level for this experiment and that the signal-to-noise ratio for the remaining steady-state backgrounds is 1:1. In Fig.~\ref{fig:sterilenu}, we calculate the rate-only 90\% confidence limits on the $\nu_\mu \rightarrow \nu_S$ mixing parameter $\sin^22\theta_{\mu S}$ for a 5-year run with C-PAR, assuming a 9\% normalization systematic uncertainty correlated between the two detectors and a 36~cm path length smearing. Also shown are the 90\% confidence limits for $\nu_\mu$ disappearance, $\nu_e$ disappearance, and $\nu_e$ appearance, assuming the $\bar\nu_\mu$ and $\nu_e$ can be detected with similar assumptions as for the $\nu_\mu$. With these parameters, PIP2-BD can definitively rule out the 3+1 global fit allowed region.

\subsubsection{Precision tests of the Standard Model}

\begin{figure}
\centering
  \includegraphics[width=0.80\textwidth]{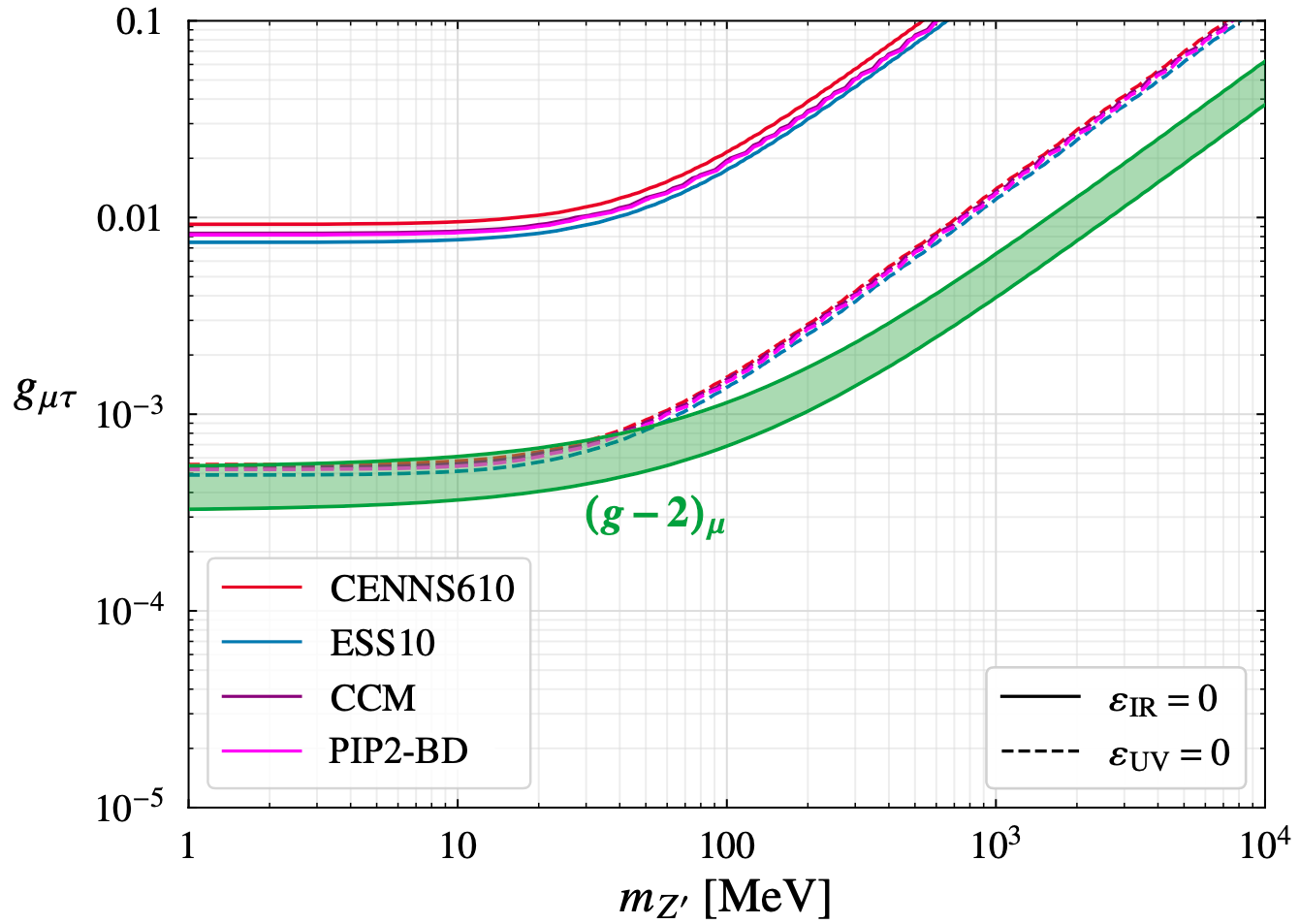}
  \caption{95\% exclusion limits for future CEvNS experiments to the $U(1)_{L_\mu-L_\tau}$ gauge symmetry scenario computed following the prescription described in Ref.~\cite{Hapitas:2021ilr}.}
  \label{fig:Lmu-Ltau}
\end{figure}

The CEvNS cross section within the SM is known to a theoretical uncertainty of 2\%.  Therefore, the measured CEvNS rate serves as a precision test of the SM and can be used to set limits on any models that modify the expected SM cross section. For example, the COHERENT experiment has used this fact to place limits on neutrino non-standard interactions~\cite{COHERENT:2017ipa,COHERENT:2019iyj,COHERENT:2020iec,Akimov:2021dab}. Here, we highlight a connection to the $(g-2)_\mu$ anomaly by considering an anomaly-free extension to the SM that introduces a $U(1)_{L_\mu-L_\tau}$ gauge symmetry.  The new heavy gauge boson kinetically mixes with the SM photon and also couples to the $\mu$- and $\tau$-flavored leptons. Such a model can explain the $(g-2)_\mu$ anomaly and would also predict a modification of the SM CEvNS cross section that can be tested by PIP2-BD~\cite{Hapitas:2021ilr}. Fig.~\ref{fig:Lmu-Ltau} shows the 95\% exclusion limits for future CEvNS experiments, assuming a normalization systematic uncertainty of 5\% and a signal-to-background ratio of 10.  The PIP2-BD limit is systematics-limited, so reducing the normalization uncertainty on the CEvNS rate will allow the setup to begin to probe the preferred parameter region for this model that explains the $(g-2)_\mu$ anomaly.

\subsubsection{Neutrino-nucleus Cross Sections}

One of the primary DUNE science goals is to study neutrinos from a core-collapse supernova burst event in detail. One of the inputs to these studies is the charged current neutrino-Ar cross section for low energy $\nu_e$, which has not been measured and for which there are large theoretical uncertainties.  Pion decay-at-rest neutrino sources such as proton beam dumps are ideal for measuring these cross sections as they provide a well-understood spectrum of $\nu_\mu$, $\bar\nu_\mu$, and $\nu_e$ neutrinos with energies up to 52.8 MeV, covering the energy range of interest. PIP2-BD will be able to perform direct measurements of the $\nu_e$ charged current cross section on argon that will improve the precision of future studies of supernova burst neutrinos in DUNE.

\section{Conclusion}

The quest to identify and understand dark matter is one of the most compelling missions in particle physics today.  As the scientific case for sub-GeV DM continues to grow, there is a clear opportunity to take advantage of the high-power proton beams that will be available at Fermilab over the coming decades. We have studied different accumulator ring scenarios that can be staged as enhancements to Fermilab's PIP-II linac to mount a HEP-dedicated beam dump facility, PIP2-BD, with world-leading sensitivity to dark sector physics and sterile neutrinos. An initial program with a 100 kW PIP-II accumulator ring could be mounted within the decade and probe new parameter space potentially explaining the dark matter relic abundance in the universe.

\bibliographystyle{unsrt}
\bibliography{bibliography}

\end{document}